\documentclass[aps,twocolumn,showpacs,superscriptaddress,citeautoscript,pra,floatfix]{revtex4-2}

\usepackage[caption=false]{subfig} 
\usepackage{ragged2e} 
\DeclareCaptionJustification{justified}{\justifying}
\usepackage[utf8]{inputenc}
\usepackage{lipsum}
\usepackage{bm}
\usepackage{graphicx,float}
\usepackage{amsmath,amssymb}
\usepackage{amsfonts}
\usepackage{color}
\usepackage{comment}
\usepackage{physics}
\usepackage{amsfonts}
\usepackage{amsmath}
\usepackage{amssymb}
\usepackage{amsthm}
\usepackage{fontenc}
\usepackage{graphicx}
\usepackage{xcolor}
\usepackage{textcomp}
\usepackage{epstopdf}
\usepackage{braket}
\usepackage{mathtools}
\usepackage{amsmath}
\usepackage{dcolumn}
\usepackage{multirow}
\usepackage{units}
\usepackage{url}

\usepackage[colorlinks=true,%
            linkcolor=blue,%
            urlcolor=blue,%
            citecolor=blue,%
            filecolor=blue,%
            bookmarksopen=true,%
            pdfauthor={DZ_JPRA_PAO},%
            pdftitle={MBS_BIC_MQDS},%
            pdfsubject={MBS_BIC_MQDS_Manuscript},%
            pdfpagemode=UseOutlines]{hyperref}
        
\definecolor{Blue}{rgb}{0.3,0.3,0.9}
\definecolor{Red}{rgb}{0.9,0.3,0.3}
\definecolor{Green}{rgb}{0.3,0.6,0.3}
\definecolor{Black}{rgb}{0.0,0.0,0.0}

\begin{document}

\title{Fano-Andreev effect in a T-shaped Double Quantum Dot in the Coulomb-blockade regime}

\author{A. Gonz\'alez I.}
\affiliation{Departamento de F\'{\i}sica, Universidad T\'{e}cnica Federico Santa Mar\'{\i}a, Casilla 110 V, Valpara\'{\i}so, Chile}
\author{A. M. Calle}
\affiliation{Departamento de F\'{\i}sica, Universidad Cat\'olica del Norte, Antofagasta, Chile}
\author{M. Pacheco}
\affiliation{Departamento de F\'{\i}sica, Universidad T\'{e}cnica Federico Santa Mar\'{\i}a, Casilla 110 V, Valpara\'{\i}so, Chile}
\author{E. C. Siqueira}
\affiliation{Departamento Acad\^emico de Física, Universidade Tecnol\'ogica Federal do Paran\'a, Brazil}

\author{Pedro A. Orellana}
\affiliation{Departamento de F\'{\i}sica, Universidad T\'{e}cnica Federico Santa Mar\'{\i}a, Casilla 110 V, Valpara\'{\i}so, Chile}



\begin{abstract}

We studied the effects of superconducting quantum correlations in a system consisting of two quantum dots, two normal leads, and a superconductor. Using the non-equilibrium Green's functions method, we analyzed the transmission, density of states, and differential conductance of electrons between the normal leads. We found that the superconducting correlations resulted in Fano-Andreev interference, which is characterized by two anti-resonance line shapes in all of these quantities. This behavior was observed in both equilibrium and non-equilibrium regimes and persisted even when Coulomb correlations were taken into account using the Hubbard-I approximation. It is worth noting that the robustness of this behavior against these conditions has not been studied previously in the literature.

\end{abstract}


\maketitle
%
%

\section{Introduction}

The investigation of hybrid structures, where normal conductors are connected to superconductors, has garnered significant interest due to their potential for applications in electronics, spintronics, and quantum information processing \cite{L.P.Kouwenhoven}. When a normal metal is connected to a superconductor lead, the superconducting order can leak into the normal metal. This causes pairing correlations and a superconducting gap due to the proximity effect\cite{buzdin2005}. The mechanism responsible for the proximity effect is known as Andreev reflection\cite{AFAndreev}.  In this process, an electron is reflected as a hole at the interface between the leads. The missing charge in the normal lead appears as a Cooper pair within the superconductor. Since the electrons in the superconductor side and the hole in the normal lead are correlated, these are represented by bound states. In effect, the two charges coming from the normal lead cannot penetrate deep into the superconductor side being absorbed into the superconductor condensate. In hybrid systems composed of quantum dots (QDs), these bound states, called Andreev bound states (ABS), appear as resonances in the QD transmission spectrum, with energies within the superconductor gap. The presence of ABSs is the key ingredient to many different features exhibited by QD-based systems\cite{Qing-feng-Sun, ZYu, Xiao-Qi-Wang, J.Gramich, Pillet, TanujChamoli, Dirks, Rosario-Fazio,JanBaranski,JanBaranski2,T.Domanski2,G.Gorski,Domanski2}. The ABSs modify the so-called Fano effect\cite{U.Fano,Miroshnichenko}, a well-known phenomenon resulting from quantum interference between discrete and continuum states. In QDs-based systems, the Fano effect signature is an asymmetric resonance pattern arising in the transmission spectrum of the QD or double quantum dots (DQDs).  Several authors have studied the effect of quantum decoherence on Fano lineshapes. To this end, they introduced a normal floating lead directly coupled to the double quantum dot (DQD). Their findings show that the floating lead coupled to the lateral QD plays a crucial role in destroying the Fano lineshape. \cite{Gao-Wen-Zhu, J.Baranski}. Other authors have studied T-shaped QDs structure coupled to two normal or ferromagnetic leads left and right and a superconducting lead \cite{Michalek,Michalek2,Yu-Zhu,G.Michalek2, E.C.Siqueira1, E.C.Siqueira2}. In particular, A. M. Calle et al. \cite{CalleAM} have shown that, in a non-interacting T-shaped double quantum dot coupled to two normal metals, the transmission between the normal leads (ET) exhibits Fano resonances due to the appearance of ABSs in the non-interacting DQD, due to the presence of the superconducting lead (Fano-Andreev effect). By using a mean-field treatment of Coulomb correlations, E. C. Siqueira \textit{et al.}. \cite{Siqueira3} have shown complementary patterns of resonances between AR and ET transmittance when a superconductor lead is coupled to a QD, which itself is coupled to two ferromagnetic leads. The effect was shown to be a result of the interplay between the ABSs and the spin polarization provided by the ferromagnets. When the charging energy is much larger than the thermal energy of the charge carriers and the QD is only weakly coupled to the leads, Coulomb correlations become increasingly significant at low temperatures. This is known as the Coulomb blockade regime, where the number of electrons in the dot is fixed. Sequential tunneling is suppressed, and transport is only possible when the energy of the state $|N_i>|$ is aligned with the chemical potential of the leads. As a result, the linear conductance through the quantum dot shows a series of peaks that correspond to the degeneracy regions (Coulomb peaks). These peaks are separated by regions of low conductance. However, when the temperature is lowered, the coupling of the electron within the QD to the electrons of the leads gives rise to virtual quantum states within the QD, which allows for an additional transport channel through the QD. This manifests as a narrow resonance peak in the transmission spectrum of the QD, called Kondo resonance. On the other hand, when one of the leads is a superconductor in a three-terminal system, it is possible to probe the interplay between the Fano effect, ET transmission, and ABS states. The interplay between ABS and Kondo effect in hybrid superconductor nanostructures has been extensively studied\cite{T.Domanski2,Grzegorz-Michalek,Sachin-Verma}. For the T-shape DQDs structure, A. M. Calle \textit{et al.}.\cite{CalleA} found that the Kondo resonance modifies the ABS resonances. However,in systems composed of DQDs, the interplay among the  Coulomb correlation within an intermediate range of values, Fano effect and ABS states, has not been studied in depth in the literature despite the potential for original effects that may be obtained in these systems.

In this work, we investigate the electronic transport properties of a T-shaped DQD system. Fig. 1 shows the scheme, where the central QD, QD$_{a}$, is connected to the normal leads while the other QD, QD$_{b}$, is coupled to a superconductor lead (S).In our analysis, we focus on the Coulomb blockade regime, with identical values for the onsite Coulomb interaction parameters in both quantum dots (QDs), namely $U_a = U_b$. To obtain our results, we use the equation-of-motion (EOM) method  to calculate the relevant Green's functions. The intradot Coulomb correlation is considered in the Hubbard-I approximation, which provides a reliable description of the Coulomb blockade regime. We investigate  in the non-equilibrium regime and at zero temperature, the impact of the Coulomb interaction (U), inter-dot coupling and coupling between QD and leads on the Fano-Andreev effect,in both molecular and interferometric regimes. To study these effects, we employ the non-equilibrium Green's function formalism.

This paper is organized as follows: in Sec. II, we present the model and formulation for the system displayed in Fig. 1,  in Sec. III, the numerical results are presented and discussed. Finally, a summary and the main conclusions are presented in Sec. IV.


\begin{figure}[ht]
\centering
\includegraphics[width= 8 cm,height=10cm, scale= 1.0]{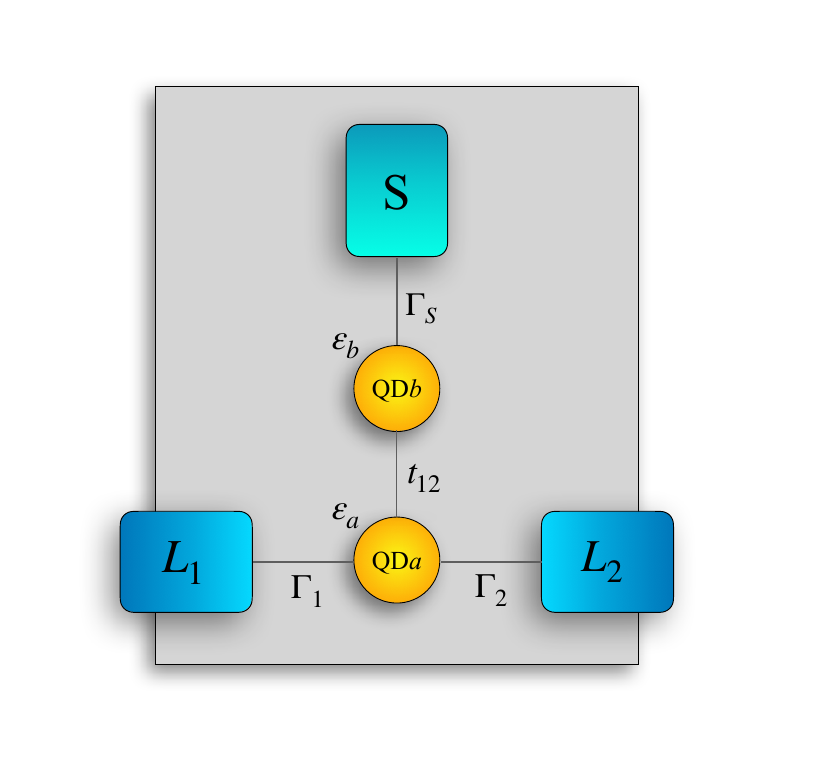}
\caption{\small{The T-shape double QD system studied in this work. It is composed of two quantum dots in which the $QD_{a}$ is coupled to two normal metals $L_{1}$ and $L_{2}$, with the coupling strength being determined by the parameters $\Gamma_{1}$ and $\Gamma_{2}$.  The second quantum dot, $QD_{b}$, is coupled to the superconductor $S$, with the coupling strength being modeled by the parameter $\Gamma_{S}$. The coupling between the QDs is modeled by the $t_{12}$ parameter. }}
\label{Fig1}
\end{figure}

\section{Model and Formulation}\label{secmodel}


In Fig. \ref{Fig1}, the T-shape double quantum-dot system is illustrated. It consists of a
central quantum dot, $QD_a$, coupled to the two normal leads, $L_{1}$ and $L_{2}$, and a side $QD_b$, connected to a superconductor $S$. The Hamiltonian of the system is given by:
\begin{equation}\label{Hamiltonian}
H = H_{1} + H_{2}+  H_{S}  + H_{DQD} + H_{T}. 
\end{equation}

The first and the second terms are the Hamiltonians for the normal leads at the left (1) and right (2) sides of $QD_{a}$. These are modeled by Eqs. \eqref{eqH1} and \eqref{eqH2}:
\begin{equation}\label{eqH1}
 H_{1}  =  \sum_{k} \sum_{\sigma} \epsilon_{1 k \sigma} c^{\dagger}_{1 k  \sigma} c_{ 1 k  \sigma},
\end{equation}

and 

\begin{equation}\label{eqH2}
 H_{2}  =  \sum_{k} \sum_{\sigma} \epsilon_{2 k  \sigma} c^{\dagger}_{2 k  \sigma} c_{2 k  \sigma}, 
\end{equation}
where $c_{  \alpha k \sigma}^{\dag}$ ($c_{  \alpha k  \sigma}$) is the electron creation (annihilation) operator of an electron with spin $\sigma$ and energy $\epsilon_{k \sigma}$ in the $\alpha$ electrode.

The second term stands for the BCS Hamiltonian\cite{BCS} of the superconducting ($S$) lead and reads:

\begin{equation}
H_{S}  =  \sum_{k}  \sum_{\sigma} \epsilon_{kS} c^{\dagger}_{k S  \sigma} c_{k S \sigma}
+ \sum_{k} (\Delta^{\ast} c_{k S \downarrow}c_{-k S \uparrow} + \Delta c^{\dagger}_{-k S \uparrow}  c^{\dagger}_{k S \downarrow} ),
\end{equation}
where $c_{k S \sigma}^{\dag}$ ($c_{k S \sigma}$) is the electron creation (annihilation) operator of an electron with spin $\sigma$ and energy $\epsilon_{kS}$ in the superconducting electrode, and $\Delta$ denotes pair potential, whose absolute value gives the superconducting energy gap. 

The  $H_{DQD}$ term is given by Eq. \eqref{HDQDeq} and takes into account the coupling between the QDs, modeled by the variable $t_{12}=t$, and the Coulomb interaction at each QD, whose strength is modeled by $U_{m}$, $m=a,b$. The QDs are assumed to have a single spin degenerated level, whose value is determined by $\epsilon_{d m, \sigma}$, $m=a,b$.


\begin{multline}
 H_{DQD} =  \sum_{m=a,b, \sigma}  \epsilon_{d m, \sigma} d^{\dagger}_{m, \sigma} d_{m, \sigma} \\
  +    \sum_{\sigma}   t  \left[ d^{\dagger}_{a \sigma} d_{b \sigma} 
  + d^{\dagger}_{a \sigma} d_{b \sigma}  \right]\\
 + \sum_{m = a,b} U_m  n_{m \sigma} n_{m \bar{\sigma}}\\
 \label{HDQDeq}
\end{multline}

\subsection{Green’s functions}

In order to obtain the transport properties for the system modeled by Eq. \eqref{Hamiltonian}, we have used the well-known non-equilibrium Green's function approach\cite{haug1996,rammer} along with the equation-of-motion method. This formalism has been extensively applied to nanostructured systems. The presence of the superconductor has been taken into account by expressing Green's functions within the Nambu space. In this way, Green's functions are represented by $4\times4$ matrices resulting from the tensor product between spin and electron-hole subspaces. By using this formalism, we have obtained a system of coupled Dyson equations for the retarded Green's functions matrices $\mathbf{G}^r_{aa}$ and $\mathbf{G}^r_{bb}$, for $QD_{a}$ and  $QD_{b}$, respectively:
\begin{equation}\label{eqGaar}
{\bf G}^r_{aa}= \mathbf{G}^{ro}_{aa}  +  {\bf G}^r_{aa} {\bf t}_{ab}^{\dagger}  {\bf G}^{ro}_{bb}    {\bf t}_{ab} {\bf G}^{ro}_{aa},
\end{equation}
and
\begin{equation}\label{eqGbbr}
{\bf G}^r_{bb}= \mathbf{G}^{ro}_{bb}  +  {\bf G}^r_{bb} {\bf t}_{ab}^{\dagger}  {\bf G}^{ro}_{aa}    {\bf t}_{ab} {\bf G}^{ro}_{bb},
\end{equation}
with 
\begin{equation}\label{eqGaar0}
    {\bf G}^{ro}_{aa}  =  {\bf g}^r_{aa}(1 - \boldsymbol{\Sigma}^{r}_L {\bf g}^r_{aa})^{-1},
\end{equation}
and 
\begin{equation}\label{eqGbbr0}
    {\bf G}^{ro}_{bb}  =  {\bf g}^r_{bb}(1 - \boldsymbol{\Sigma}^{r}_S {\bf g}^r_{bb})^{-1}.
\end{equation}

In Eqs. \eqref{eqGaar} to \eqref{eqGbbr0}, $ {\bf g}^r_{aa}$ and $ {\bf g}^r_{bb}$ are the $QD_{a}$ and $QD_{b}$ Green’s functions, respectively, when the QDs are isolated from the leads; $\mathbf{t}_{ab}$ describes the tunneling between $QD_{a}$ and $QD_{b}$. The coupling to the normal and superconducting leads are modeled by retarded/advanced self-energies $\boldsymbol{\Sigma}^r_L = \boldsymbol{\Sigma}^r_{1} + \boldsymbol{\Sigma}^r_{2}$ and $\boldsymbol{\Sigma}^r_S$, respectively.

The self-energy for the coupling to the normal leads ($L_{1}$ and $L_{2}$) is given by Eq. \eqref{SigmaL}:
\begin{equation}\label{SigmaL}
\boldsymbol{\Sigma}_{L}^{r,a}= \mp \dfrac{i}{2}(\Gamma_{1}+\Gamma_{2})
\begin{bmatrix}
 1&0&0&0
 \\
 0&1&0&0
 \\
 0&0&1&0
 \\
 0&0&0&1
\end{bmatrix},
\end{equation}
where $\Gamma_{i}=\Gamma_{i\uparrow}+\Gamma_{i\downarrow}$ and $\Gamma_{i\sigma} = 2 \pi |t_{i}|^2 N_{i} $, ($i =  1, 2$) being the coupling strength, with $t_i$ being amplitude for an electron with spin $\sigma$ of $QD_{a}$ to be transferred to the lead $L_{i}$; $N_{i}$ is the density of states at Fermi level for the normal lead, $L_{i}$. Since we have assumed both normal leads as non-magnetic, the density of states is the same for both spins.

The retarded/advanced self-energy of the superconductor is given by Eq. \eqref{Sigmas}:
\begin{equation}\label{Sigmas}
{\boldsymbol \Sigma}^{r,a}_{S}=  \mp \frac{i}{2} \rho_S(\epsilon) \Gamma_S  \begin{bmatrix}
  1  &    - \dfrac{\Delta}{\epsilon} & 0 & 0    \\ 
 - \dfrac{\Delta}{\epsilon}  &  1  & 0 & 0    \\
0 & 0 &  1  &    \dfrac{\Delta}{\epsilon}    \\ 
0 & 0 &   \dfrac{\Delta}{\epsilon}  &  1  \\
  \end{bmatrix}.
\end{equation}

In Eq. \eqref{Sigmas}, $\Gamma_S=2\pi|t_{s}|^{2}N_{s}$ is the coupling strength between the superconductor lead and $QD_{b}$, defined in terms of the amplitude of tunneling $ |t_{s}|$  and the normal density of states of $N_{s}$. The $\Delta$ appearing in some of the matrix elements stands for the energy gap of the superconductor and accounts for the electron-hole coupling. The energy gap plays a central role in this model and also modifies the self-energy through $\rho_S$, the dimensionless modified BCS density of states, whose expression is given by:
\begin{equation}
\rho_{S}(\epsilon) = \frac{| \epsilon | \theta(| \epsilon | - \Delta)}{\sqrt{ \epsilon^2 - \Delta^2}} 
- i \frac{\epsilon  \theta(\Delta - | \epsilon | )}{\sqrt{ \Delta^2 - \epsilon^2 }},
\end{equation}
with the imaginary part accounting for the Andreev bound states (ABS), within the superconductor gap.

The presence of electronic correlations associates the QD energy levels with the electronic occupations, which, in turn, depend on external parameters like gate and bias voltages. As a result, Eqs. \eqref{eqGaar} and \eqref{eqGbbr} must be solved in a self-consistent way, together with the occupations of the QDs. Such occupations are obtained from the diagonal matrix elements of the ``lesser'' Green’s function matrix, which is obtained through the Keldysh equation. For the $QD_{a}$, the expression reads:
\begin{equation}\label{Glessera}
 { \bf G}^{<}_{aa}= { \mathbf G}^{r}_{aa}(\omega) {\boldsymbol \Sigma}^{<}_{Ta} {\bf G}^{a}_{aa}(\omega),
\end{equation}
with
\begin{equation}
 {\boldsymbol \Sigma}^{<}_{Ta} ={\boldsymbol \Sigma} ^{<}_{L}(\epsilon)  + {\bf t}_{ab}^{\dagger} {\bf G}^{ro}_{bb} {\boldsymbol \Sigma}^{<}_{S}(\epsilon) {\bf G}^{ao}_{bb} {\bf t}_{ab}.
\end{equation}

The expression for $QD_b$ can be obtained by exchanging the indices $a$ and $b$. In Eq. \eqref{Glessera}, ${\boldsymbol \Sigma}^{<}_{Ta}$ represents the ``lesser'' self-energy, which is expressed in terms of the self-energies of the leads: $\boldsymbol{\Sigma}^{<}_{L}=\boldsymbol{\Sigma}^{<}_{1} + \boldsymbol{\Sigma}^{<}_{2}$ and $\boldsymbol{\Sigma}^{<}_{S}$. Assuming that the leads are in equilibrium with well-defined chemical potential and temperature, the self-energies of the leads can be obtained using the fluctuation-dissipation theorem: $\boldsymbol{\Sigma}^{<}_i= {\bf F}_{i} \left[ \boldsymbol{\Sigma}^a_i - \boldsymbol{\Sigma}^r_i \right]$, where the Fermi matrix $\mathbf{F}_{i}$ is given by

\begin{equation}
{\bf F}_i  =  i \begin{bmatrix} 
 f_{i}  &  0  & 0 & 0 \\ 
0  &    \bar{f}_{i}  & 0 & 0 \\      
0 & 0 & f_{i}  &  0  \\ 
0 & 0 & 0  &    \bar{f}_{i}
    \end{bmatrix},
\end{equation}
with $f_i = f(\epsilon - e V_i)$ and $ \bar{f}_i = f(\epsilon + e V_i)$ ($i=1,2,S$) being the Fermi functions for electrons and holes, respectively. Since the superconductor is assumed to be grounded, $f_i = f(\epsilon)$ for $i=s$.

\subsection{Transmittance and current}

The presence of Coulomb correlations leads to a self-consistent problem when solving Eqs. \eqref{eqGaar} and \eqref{eqGbbr}, since the dependency on the occupations wraps the Green's functions to the occupancies of the QDs. Within the equation-of-motion approach used in this work, this results in an infinite set of equations with Green's functions of increasing order of complexity. In order to obtain a closed set of equations, one needs to resort to some approximation on the Coulomb correlations. In this work, we have used the so-called Hubbard-I approximation\cite{hubbardI}, which allows us to derive a simple expression for the electrical current in terms of the different transmission amplitudes that contribute to the electronic transport for this system. In fact, the current $I_{j}$, flowing in the lead $L_{j}$ ($j=1,2$), is given by the following expression:

\begin{multline}\label{currentmatrix}
I_{j} =  \frac{ e}{\hbar}  \int \, d\epsilon  \, [   {\bf G}^{r}_{aa}(\epsilon) {\boldsymbol \Sigma}^{<}_{j}(\epsilon) + 
 {\bf G}^{<}_{aa}(\epsilon) {\boldsymbol\Sigma}^{a}_{j}(\epsilon) \\
 + \text{H.c.} ]_{(11 + 33)}, \qquad j=1, 2.\\
\end{multline}
where ${11+33}$ stands for the sum of the 11 and 33 matrix elements of the current matrix. By substituting the Green's functions and self-energies into Eq.  \eqref{currentmatrix}, one obtains the main expression for the electric current flowing in the lead $L_1$, written in terms of the transmittances:
\begin{multline}\label{currentL1}
I_{1} = \frac{e}{h} \int  [T^{DAR}_{11} (f_1 - \bar{f}_1) +  T^{ET}_{12} (f_1 - f_2) + T^{CAR}_{12} (f_1 - \bar{f}_2) \\
+ T^{QP}_{1S}(f_1 - f_S)] d\epsilon,
\end{multline}

The first term in Eqs. \eqref{currentL1} , $T^{DAR}_{11}$,  corresponds to direct Andreev reflection transmittances through the paths ($L_1-QD_a-QD_b-S$), i.e., an electron of $L_1$ is reflected by $S$ into a hole of $L_1$. The second term, $T^{ET}_{12}$, represents electron tunneling (ET) between the normal leads via ($L_1-QD_a-QD_b-L_{2}$) path. The next term, $T^{CAR}_{11}$, account for the transmittances of crossed Andreev reflection through the path ($(L_1,L_2)-QD_a-QD_b-S)$, i.e., an electron of $L_1$ is reflected by $S$ into a hole of $L_2$. Finally, the last term, $T^{QP}_{11}$, corresponds to quasiparticles tunneling through the ($L_1-QD_a-QD_b-S$) path.

The amplitudes can be expressed in terms of Green's functions, which are given as follows:
\begin{equation}
  \begin{split}
      T^{DAR}_{11}& = \Gamma_1^2 ( |G^{r}_{aa,14}|^2 + | G^{r}_{aa,12}|^2 + |G^{r}_{aa,34}|^2 + | G^{r}_{aa,32}|^2) \\
       T^{CAR}_{12}& = \Gamma_1 \Gamma_2 ( |G^{r}_{aa,14}|^2 + | G^{r}_{aa,12}|^2 + |G^{r}_{aa,34}|^2 + | G^{r}_{aa,32}|^2) \\
        T^{ET}_{12}& = \Gamma_1 \Gamma_2 ( |G^{r}_{aa,33}|^2 + | G^{r}_{aa,31}|^2 + |G^{r}_{aa,13}|^2 + | G^{r}_{aa,11}|^2) \\
        T^{QP}_{1S} & = \bar{\rho} \Gamma_1 \Gamma_S \{  Y^{-}_{21}   |G^{r}_{aa,12}|^2 
   +   X^{+}_{34}     | G^{r}_{aa,13}|^2  +  Y^{+}_{43}    |G^{r}_{aa,14}|^2 \\
   & +  X^{-}_{12}  |G^{r}_{aa,11}|^2 - Z^{+}_{34} G^{r}_{aa,13}  [G^{r}_{aa,14}]^{\ast} 
    -  [Z^{+}_{34}   ]^{\ast}  [G^{r}_{aa,13}]^{\ast} G^{r}_{aa,14}
    \\
   &  - Z^{-}_{12}     G^{r}_{aa,11}  [G^{r}_{aa,12}]^{\ast}- [Z^{-}_{12}]^{\ast}  [G^{r}_{aa,11}]^{\ast} G^{r}_{aa,12}
     \\
 & +   Y^{-}_{21}   |G^{r}_{aa,32}|^2 
   +   X^{+}_{34}     | G^{r}_{aa,33}|^2  +  Y^{+}_{43} |G^{r}_{aa,34}|^2 \\
   & +  X^{-}_{32}  |G^{r}_{aa,31}|^2 - Z^{+}_{34} G^{r}_{aa,33}  [G^{r}_{aa,34}]^{\ast} \\
   & -  [Z^{+}_{34}]^{\ast}  [G^{r}_{aa,33}]^{\ast} G^{r}_{aa,34} 
    - Z^{-}_{12}     G^{r}_{aa,31}  [G^{r}_{aa,32}]^{\ast} \\
    & - [Z^{-}_{12}]^{\ast}  [G^{r}_{aa,31}]^{\ast} G^{r}_{aa,32}
    \} \\
  \end{split}  
\end{equation}
where
\begin{equation}
  \begin{split}
       X^{\pm}_{ij} & \equiv   t^2   [   |G^{ro}_{bb,ii}|^2  +   |G^{ro}_{bb,ij}|^2  \pm   \frac{\Delta}{\epsilon} (G^{ro}_{bb,ii} [G^{ro}_{bb,ij}]^{\ast} + G^{ro}_{bb,ij} [G^{ro}_{bb,ii}]^{\ast}),\\ 
          Y^{\pm}_{ij} & \equiv   t^2   [   |G^{ro}_{bb,ii}|^2 +   |G^{ro}_{bb,ji}|^2  \pm   \frac{\Delta}{\epsilon} (G^{ro}_{bb,ii} [G^{ro}_{bb,ji}]^{\ast} + G^{ro}_{bb,ji} [G^{ro}_{bb,ii}]^{\ast}) ],\\
          Z^{\pm}_{ij} & \equiv   t^2   [ G^{ro}_{bb,ij} [G^{ro}_{bb,jj}]^{\ast}
            + [G^{ro}_{bb,ij}]^{\ast}  G^{ro}_{bb,ii} \\
             &\qquad      \qquad  \qquad   \pm        \frac{\Delta}{\epsilon} (|G^{ro}_{bb,ij}|^{2} + [G^{ro}_{bb,jj}]^{\ast} G^{ro}_{bb,ii}) ],\\
  \end{split}
\end{equation}

(The current flowing and transmittances in $L_{2}$ can be obtained by interchanging the indexes 1 and 2 in the above equation).

\subsection{Self-consistent calculations}
Regarding the occupations of the QDs, they are determined by the ``lesser'' Green's function. In the case of normal leads with no polarization, the average occupation number remains independent of spin. This allows us to set, for each QD, $\langle n_{i, \sigma} \rangle = \langle n_{i} \rangle$ where $i = a, b$. These occupation numbers are obtained by solving the following self-consistent system of integral equations: \begin{subequations}\label{iteracoes}
\begin{align}
\langle n_{a \uparrow} \rangle &= - i \int \frac{d \epsilon}{2 \pi} \, G^{<}_{aa,11}[\epsilon, \langle n_{a \uparrow}\rangle,  \langle n_{a\downarrow}\rangle , \langle n_{b \uparrow}\rangle, \langle n_{b \downarrow}\rangle],
\\ \langle n_{a \downarrow} \rangle &= - i \int \frac{d \epsilon}{2 \pi} \, G^{<}_{aa,33}[\epsilon, \langle n_{a \uparrow}\rangle,  \langle n_{a\downarrow}\rangle , \langle n_{b \uparrow}\rangle, \langle n_{b \downarrow}\rangle],\\
\langle n_{b \uparrow} \rangle &= - i \int \frac{d \epsilon}{2 \pi} \, G^{<}_{bb,11}[\epsilon, \langle n_{a \uparrow}\rangle,  \langle n_{a\downarrow}\rangle , \langle n_{b \uparrow}\rangle, \langle n_{b \downarrow}\rangle],\\
\langle n_{b \downarrow} \rangle &= - i \int \frac{d \epsilon}{2 \pi} \, G^{<}_{bb,33}[\epsilon, \langle n_{a \uparrow}\rangle,  \langle n_{a\downarrow}\rangle , \langle n_{b \uparrow}\rangle, \langle n_{b \downarrow}\rangle]. 
\end{align}
\end{subequations}

In the electron-hole symmetry point, i.e., when $\epsilon_d = -U/2$, the occupation is independent of bias voltage and equal to $1/2$ ($n_{a,\sigma}  = n_{b,\sigma} = 0.5$, with $\sigma = \uparrow, \downarrow$).  In this way, the exact expression for  $(dI/dV)^{ET}$ and $(dI/dV)^{AR}$ in function of bias voltage can be calculated analytically, and is equal to:

\begin{equation}
\frac{dI^{ET}}{dV} = 2 \frac{(4 G\,A - 2 F\,B \Gamma_S^2 + \frac{1}{4} C \Gamma_S^4)}{(K)} 
\end{equation}
and 
\begin{equation}
\frac{dI^{AR}}{dV} = 4 \frac{t^4 \,  \Gamma_S^2}{(K)} 
\end{equation}

where 

\begin{multline}
    K = 4  D \, A - 
  2 (-t^4 \Gamma_L^2 + G\,F\,B \\ 
  -  F (t^2 - \Gamma_L^2)  B) \Gamma_S^2 + 
  \frac{1}{4} E\,C \Gamma_S^4
\end{multline}
with
\begin{align}
  A &= \left[(\bar{g}^r_{1,12})^{2} -t^2\right]^2 + (\bar{g}^r_{1,12})^{2} \Gamma_L \nonumber \\
  B &= (\bar{g}^r_{1,12})^{2} - t^2 + \Gamma_L^2 \nonumber \\
  C &=  (\bar{g}^r_{1,12})^{2} + \Gamma_L^2 \nonumber \\
  D &= [(\bar{g}^r_{1,11})^{2} - t^2]^2 + 
  (\bar{g}^r_{1,11})^{2} \, \Gamma_L^2 \nonumber \\
  E &=  (\bar{g}^r_{1,11})^{2} + \Gamma_L^2 \nonumber \\
  F &= \bar{g}^r_{1,11} \bar{g}^r_{1,12} \nonumber \\
  G  &= (\bar{g}^r_{1,11})^{2} \nonumber \\
\end{align}
while  $\bar{g}^r_{1,11}$ and $\bar{g}^r_{1,12}$ are the 
components $11$ and $12$ of the inverse Green function of $QD_1$ isolated in the Hubbard-I approximation, evaluated in the bias voltage $V$, i.e.:

\begin{align}
  \bar{g}^r_{i,11} &= \frac{((V - \epsilon_d - U) (V - \epsilon_d))}{(V - \epsilon_d + U (n_1 - 1))}  \\
  \bar{g}^r_{i,12} &= \frac{((V + \epsilon_d + U) (V + \epsilon_d))}{(V + \epsilon_d - U (n_1 - 1))}\\
  \end{align}

The analysis of these curves is complemented by the local density of states of the QDs, $\rho_{a}$ and $\rho_{b}$, defined according to:
\begin{equation}\label{ldos}
\rho(\epsilon)_{i}=-\dfrac{1}{\pi}\text{Im}(\mathbf{G}^{r}_{ii,11}+\mathbf{G}^{r}_{ii,33}), \qquad i=a, b.
\end{equation}

\section{Results and Discussion}

In what follows, we consider $\Gamma_{1}$ as the energy unit. We assume that the QDs levels are spin degenerate and that the intradot Coulomb interaction is the same in both QDs, i.e., $U_a=U_b=U$. We denoted $r$ as the ratio of leads coupling $\Gamma_S/\Gamma_1$ through which we control the efficiency of the proximity effect, as well as the interdot tunneling as $t$.   Also, we set the gate voltage for each QD in the electron-hole symmetry point, i.e., $\epsilon_a = \epsilon_b = -U/2$. In our work, we focus on the non-equilibrium regime, and the analysis of the results is split into two parts: interferometric regime (when $t < \Gamma_1$ )  and molecular regime  (when $t \geq \Gamma_1$ ). In our analysis, we present the results for the differential conductance ET ( $(dI/dV)^{ET}_{12}$ ) and the differential conductance AR $(dI/dV)^{DAR}_{12}$ in terms of the bias voltage applied to the leads, which we will refer to for simplicity as $(dI/dV)^{ET}$  and $(dI/dV)^{DAR}$, respectively. The chemical potential of the normal leads are set with opposite bias voltage, i.e. $\mu_{1}= - \mu_{2}= eV$, while the superconductor is kept grounded, $\mu_{S}=0$. With the normal leads with opposite bias voltage, $(dI/dV)^{CAR}_{12}$ (or $I^{CAR}_{12}$)  and  $(dI/dV)^{CAR}_{21}$ (or $I^{CAR}_{21}$)   are zero since they are proportional to the factor $(f_1 - \bar{f}_2) $ and $(f_2 - \bar{f}_1)$, respectively, which are identically zero. It is worth saying that we are assuming that $|e V | < \Delta$, and therefore the contribution of the quasiparticle current  $I^{QP}$ and transmittance $T^{QP}$ is zero within the energy range we are considering. Finally, it is important to note that our calculations were obtained out of the Kondo regime.

\subsection{Interferometric regime}

\begin{figure*}[ht]
\includegraphics[width= 14 cm,height= 8 cm, scale= 1.0]
{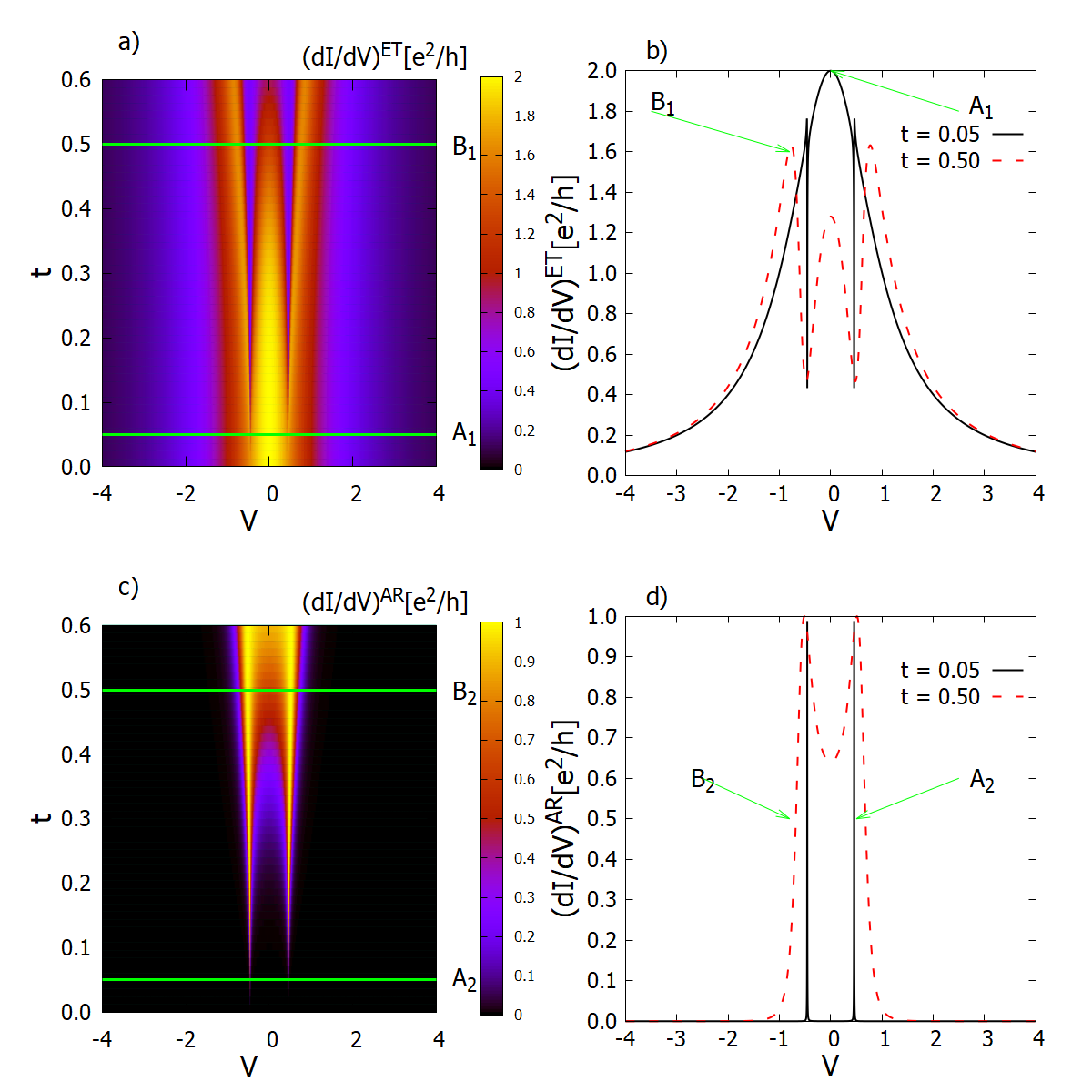}
\caption{\small{Left panel: Contour plot of 
$(dI/dV)^{ET}$ and $(dI/dV)^{AR}$ in terms  of $t$ and $V$, figures a) and c), respectively ; Right panel:$(dI/dV)^{ET}$ and $(dI/dV)^{AR}$  in terms of the energy when $t = 0.05$ (black solid line)  or $t = 0.5$(red dashed line), figures b) and d), respectively. Their location at the contour plot are indicated by the horizontal lines labeled by  A\textsubscript{1} and A\textsubscript{2}  for $t = 0.05$  and by B\textsubscript{1} and B\textsubscript{2}  for $t = 0.5$ , respectively. Other parameters are chosen as $U = 0$, $k_B T = 0 \Gamma_1$, $\Gamma_1=\Gamma_2 = \Gamma_S$ and  $\epsilon_{a} = \epsilon_{b} =-U/2$   }}
\label{figure2}
\end{figure*}


\begin{figure*}[ht]
\includegraphics[width= 14 cm,height= 8 cm, scale= 1.0]
{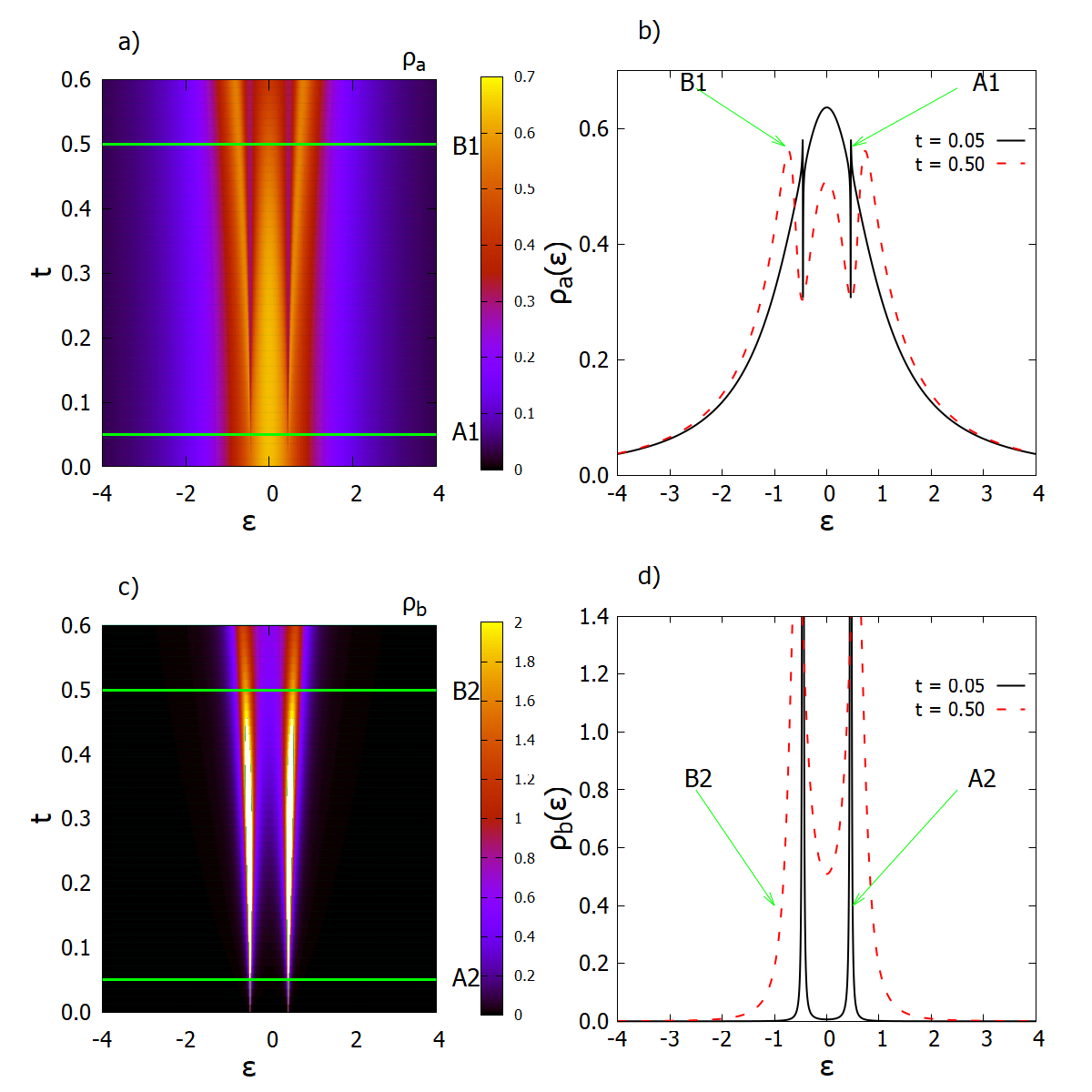}
\caption{\small{ Left panel: Contour plot of 
$\rho_a$ and $\rho_b$ in terms  of $t$ and $\epsilon$, figures a) and c), respectively ; Right panel:$\rho_a$ and $\rho_b$  in terms of the energy when $t = 0.05$(black solid line)  or $t = 0.5$(red dashed line), figures b) and d), respectively. Their location at the contour plot is indicated by the horizontal lines labeled by  A\textsubscript{1} and A\textsubscript{2}  for $t = 0.05$  and by B\textsubscript{1} and B\textsubscript{2}  for $t = 0.5$, respectively. Other parameters are chosen as $U = 0$, $k_B T = 0 \Gamma_1$, $\Gamma_1=\Gamma_2 = \Gamma_S$ and  $\epsilon_{a} = \epsilon_{b} =-U/2$  }}
\label{figure3}
\end{figure*}


\begin{figure*}[ht]
\centering
\includegraphics[width= 14 cm,height= 8 cm, scale= 1.0]
{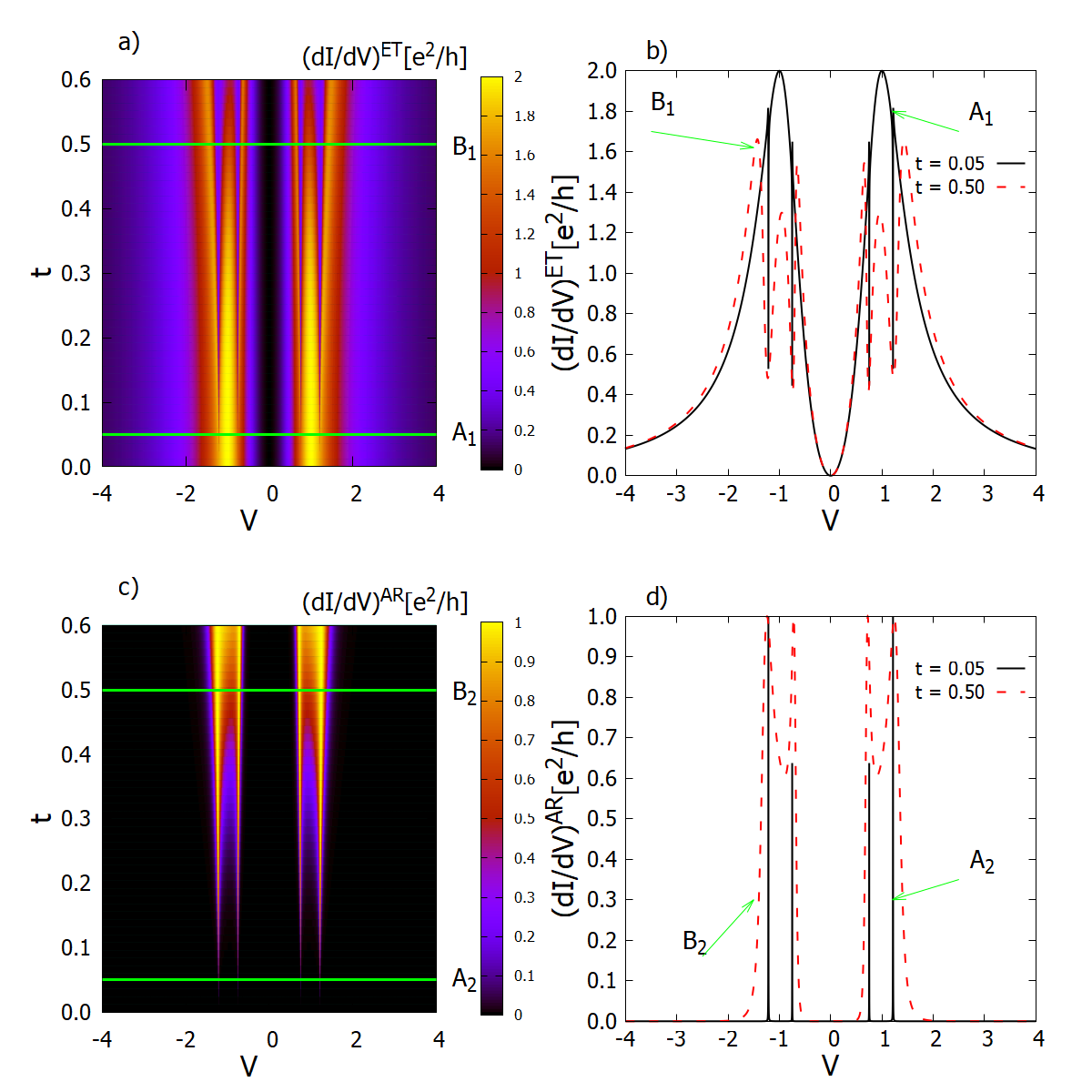}
\caption{\small{ Left panel: Contour plot of 
$(dI/dV)^{ET}$ and $(dI/dV)^{AR}$ in terms  of $t$ and $V$, figures a) and c), respectively ; Right panel:$(dI/dV)^{ET}$ and $(dI/dV)^{AR}$  in terms of the bias voltage V,  when $t = 0.05$ (solid line)  or $t = 0.5$(dashed line), figures b) and d), respectively. Their location at the contour plot is indicated by the horizontal lines labeled by  A\textsubscript{1} and A\textsubscript{2}  for $t = 0.05$  and by B\textsubscript{1} and B\textsubscript{2}  for $t = 0.5$, respectively. Other parameters are chosen as $U = 2$, $k_B T = 0 \Gamma_1$, $\Gamma_1=\Gamma_2 = \Gamma_S$ and  $\epsilon_{a} = \epsilon_{b} =-U/2$  }}
\label{figure4}
\end{figure*}

\begin{figure*}[ht]
\centering
\includegraphics[width= 14 cm,height= 8 cm, scale= 1.0]
{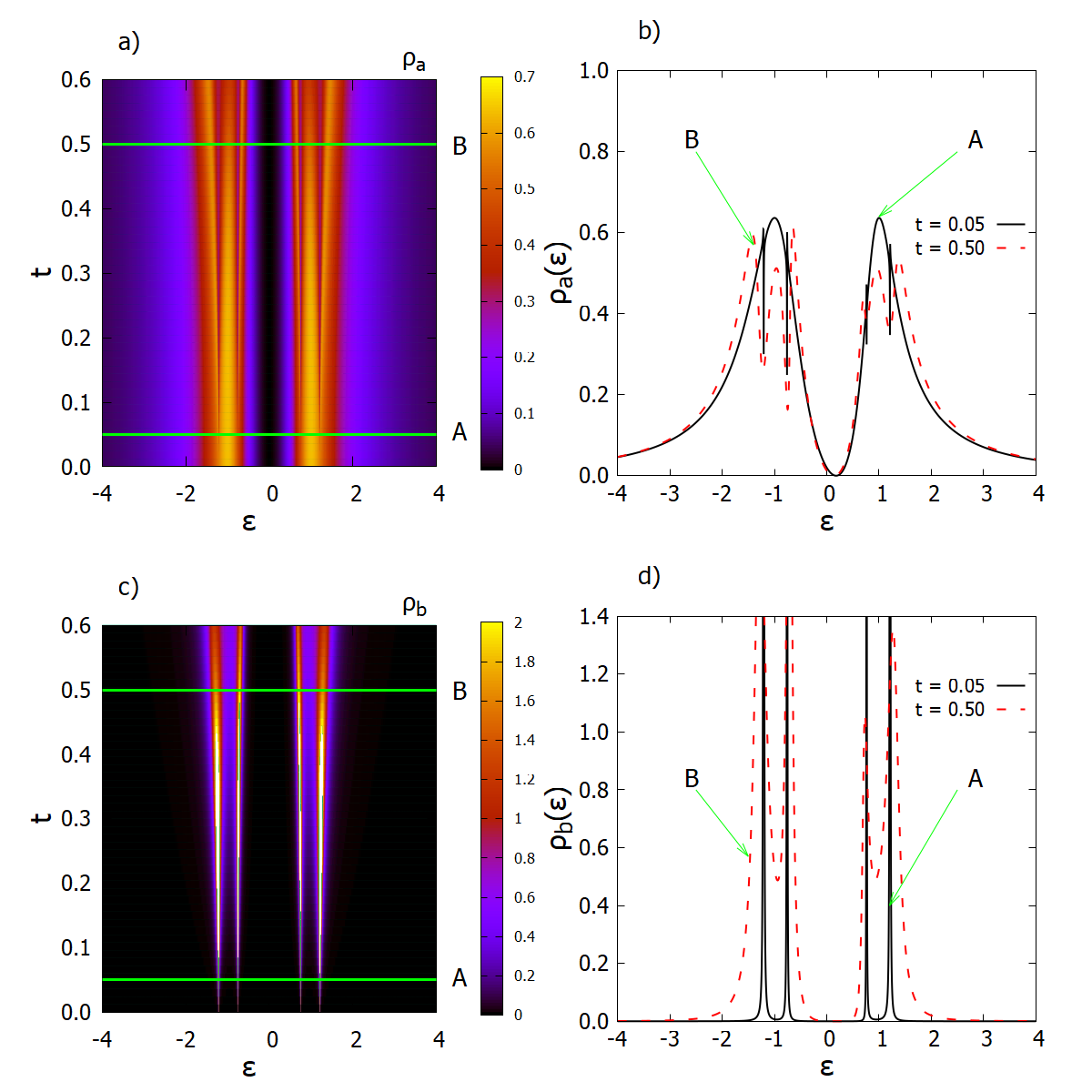}
\caption{\small{ Left panel: Contour plot of 
$\rho_a$ and $\rho_b$, at zero bias voltage,  in terms  of $t$ and $\epsilon$, figures a) and c), respectively ; Right panel:$\rho_a(\epsilon)$ and $\rho_b(\epsilon)$  in terms of the energy when $t = 0.05$(solid line)  or $t = 0.5$(dashed line), figures b) and d), respectively.Their location at the contour plot are indicated by the horizontal lines labeled by  A\textsubscript{1} and A\textsubscript{2}  for $t = 0.05$  and by B\textsubscript{1} and B\textsubscript{2}  for $t = 0.5$, respectively. Other parameters are chosen as $U = 2$, $k_B T = 0 \Gamma_1$, $\Gamma_1=\Gamma_2 = \Gamma_S$ and  $\epsilon_{a} = \epsilon_{b} =-U/2$  }}
\label{figure5}
\end{figure*}

We begin the analysis of non-equilibrium regime by focusing on the  interdot tunneling effects on $(dI/dV)^{ET}$ and $(dI/dV)^{AR}$, within the interferometric regime $(t\leq\Gamma_{1})$. For clarity's sake,  we start by focusing on the noninteracting regime ($U=0$). In this case,   the behavior of $(dI/dV)^{ET}$ and $(dI/dV)^{AR}$  as functions of $t$ and $V$, are shown in the contour plots of Figs. \ref{figure2}(a) and \ref{figure2}(c), respectively. For $t=0$, the $QD_{a}$ is decoupled from $QD_{b}-S$ part of the system and, as a result,$(dI/dV)^{ET}$ exhibits a well-known Lorentzian shape, while the $(dI/dV)^{AR}$ remains equal to zero. As $t$ increases, the superconducting correlations start to leak into $QD_{a}$, and the Andreev bound states (ABSs) emerge as two resonances equidistant from $V=0$. At the same bias voltage values, two equidistant deeps appear in the $(dI/dV)^{AR}$ curves. The correspondence between these two lineshapes can be better observed in Figs. \ref{figure2}(b) and \ref{figure2}(d) where cuts of the contour plots for $t=0.05$ ($\mathsf{A}_{1}$ and $\mathsf{A}_{2}$ lines) and  and $t=0.5$ ($\mathsf{B}_{1}$ and $\mathsf{B}_{2}$ lines) are shown. It is also worth noting that the central peak at $V = 0$ is preserved in $(dI/dV)^{ET}$ curves, as a manifestation of electron-hole symmetry of the ABSs on the QDs spectra. As $t$ increases, the hybridization between the discrete ABSs with the continuum states, stemming from the normal leads, causes the lineshapes to broaden. This interpretation is based on the fact that the differential conductance curves follow the behavior of the local density of states of the QDs for zero bias voltage, as shown in Fig. \ref{figure3}. The very same behavior can be observed from the contour plots for $\rho_{a}$  and $\rho_{b}$, at zero bias voltage, as shown in Figs. \ref{figure3}(a) and \ref{figure3}(c), respectively.

\begin{figure*}[ht]
\centering
\includegraphics[width= 14 cm,height= 8 cm, scale= 1.0]
{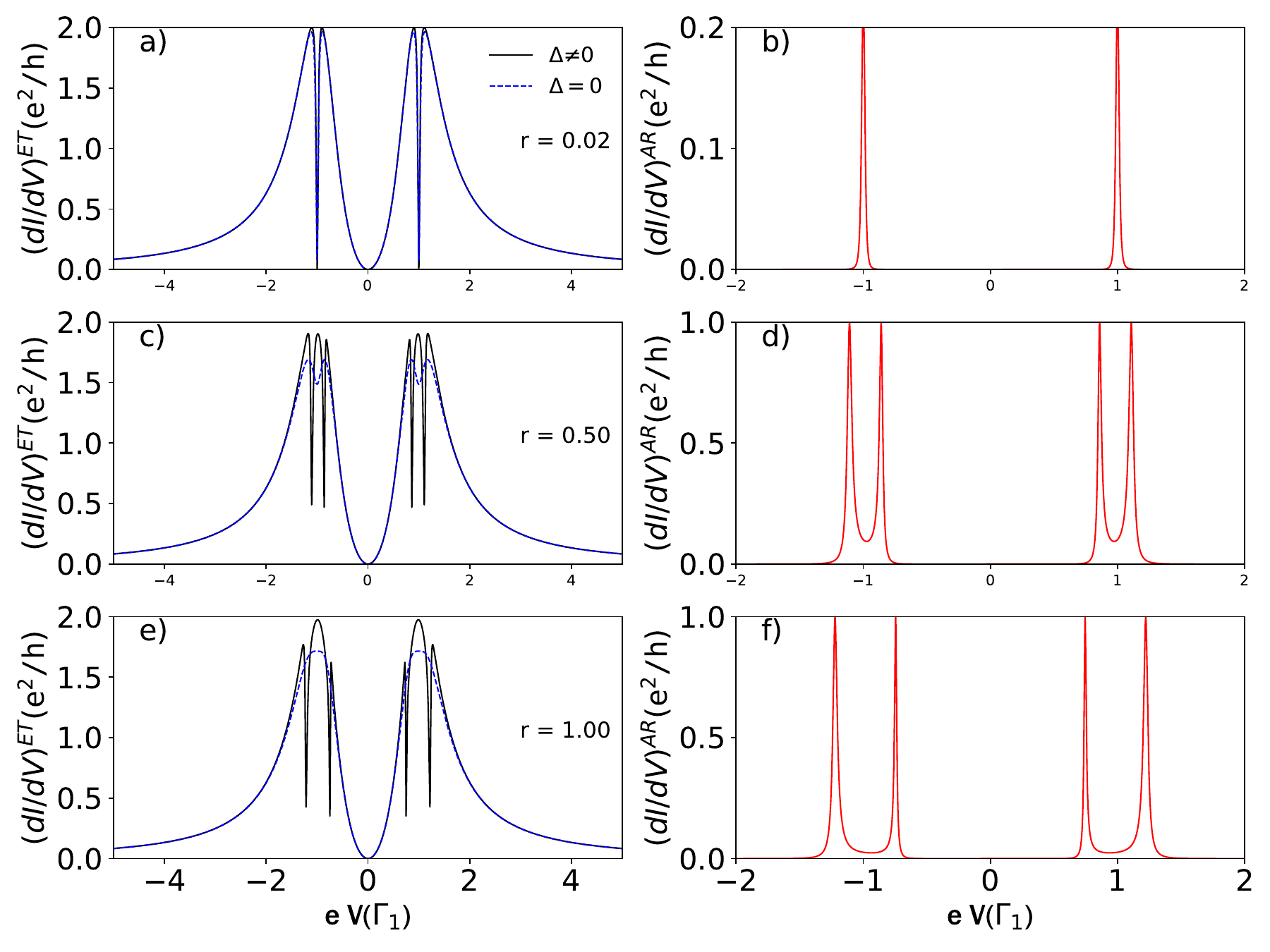}
\caption{\small{The differential conductance $(dI/dV)^{ET}$ (solid black line)  and  $(dI/dV)^{AR}$ (dashed red line) as a function of the bias voltage $V$  for a) $ r = 0.02$, b) $r = 0.50$ and c) $r = 1$. The dashed line in the left panel corresponds to the $(dI/dV)^{ET}$  for $\Delta = 0$, i.e. for a system with three normal contacts. Other parameters are chosen as:  $\Delta = 5 \Gamma_1$,   $U = 2 \Gamma_1$, $t = 0.2 \Gamma_1$, $k_B T = 0 \Gamma_1$, $\Gamma_1=\Gamma_2$  and  $\epsilon_{a} =\epsilon_{b} = -U/2$  }}
\label{figure6}
\end{figure*}

\begin{figure*}[ht]
\centering
\includegraphics[width= 14 cm,height= 8 cm, scale= 1.0]
{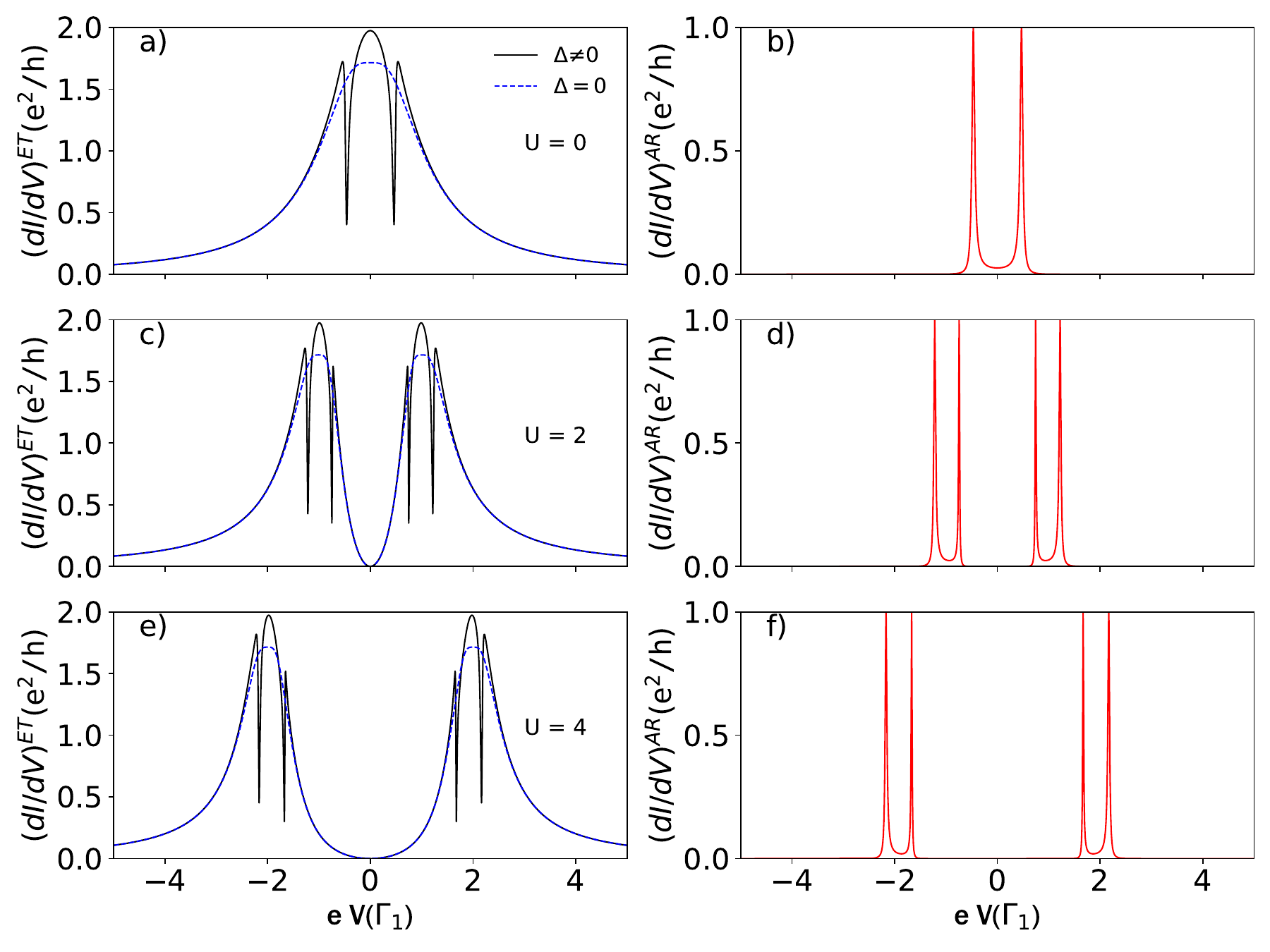}
\caption{\small{In the left panel, the differential conductance $(dI/dV)^{ET}$ (black solid line )  as a function of the bias voltage $V$ and in the right panel $(dI/dV)^{DAR}$ (red solid line ) as a function of the bias voltage $V$ when  $ U = 0$ for a) and b), $ U = 2$ for c) and d), and $ U = 4$ for e) and f). The dashed line in the left panel corresponds to the $(dI/dV)^{ET}$  for $\Delta = 0$, i.e., for a system with three normal contacts. Other parameters are chosen as $t = 0.2 \Gamma_1$ , $\Delta = 5 \Gamma_1$ , $k_B T = 0 \Gamma_1$, $\Gamma_1=\Gamma_2=\Gamma_S$  and  $\epsilon_a = \epsilon_b =-U/2$}.} 
\label{figure7}
\end{figure*}

 A more detailed lineshape can be seen in the cuts for $t=0.05$ ($\mathsf{A}_{1}$ and $\mathsf{A}_{2}$ lines) and  $t=0.5$ ($\mathsf{B}_{1}$ and $\mathsf{B}_{2}$ lines) which readily show the same behavior observed in the transmission curves.





In Fig. \ref{figure4}, the differential conductance $(dI/dV)^{ET}$ and $(dI/dV)^{AR}$  are shown for $U=2$. The other parameters remain the same as the ones used in Figs. \ref{figure2} and \ref{figure3}. The main difference, in this case, is the splitting of the peaks due to the Coulomb correlation. In fact, for $t=0.05$, one observes two Lorentzian peaks as evident from the cut $\mathsf{A}_{1}$, shown in Fig. \ref{figure4}(b). The ABSs are also evident at this small value of $t$ as anti-resonances superimposed to the peaks in spite of the presence of Coulomb correlations. As $t$ is increased, the broadening of peaks increases, as evident from the contour plots of Figs. \ref{figure4}(a) and \ref{figure4}(c) and the corresponding cuts shown Figs. \ref{figure4}(b) and \ref{figure4}(d). In contrast to the broadening of the peaks, their location is fixed by the Coulomb interaction with the two Lorentzian peaks located at $\pm U/2$, which is also the center of symmetry of the ABSs. Similar behavior is observed for the local density of states for zero bias voltage, as shown in Fig. \ref{figure5}. This figure shows in the left panel the contour plot of $\rho_a$ and  $\rho_b$ as a function of $\epsilon$ and $t$ [Figs 5 (a) and  5(c), respectively] and  shows in the right panel the curves of $\rho_a(\epsilon)$ and $\rho_b(\epsilon)$ in terms of the energy [Figs. 5 (b) and 5 (d), respectively], when $t = 0.05$ (solid line) or $t = 0.5$ (dashed line). Their location at the contour plot is indicated by the horizontal lines labeled by A\textsubscript{1} and A\textsubscript{2} for $t = 0.05$ and  B\textsubscript{1} and B\textsubscript{2} for $t = 0.5$. As we can see in figures 5 a) and 5 c), $\rho_a$  takes the form of two  Lorentzian curves centered at $\epsilon = \pm U/2$ when $t = 0$, while $\rho_b$ equals zero. However, when $t=0.05$ , we observe the appearance of two pairs of narrow deeps in $\rho_a$ [see lines  $\mathsf{A}_{1}$, in Figs. 5(a) and 5(b)] revealing ABS leakage in the density of states of   QD\textsubscript{a} , expressed in the two pairs of sharp resonances in $\rho_b$ [see lines $\mathsf{A}_{2}$ , in Figs. 5(c) and 5(d)].  However, as the value of $t$ increases, for example, when $t=0.5$ the deeps in the $\rho_a$ [see lines $\mathsf{B}_{1}$, in Figs. 5 (a) and 5 (b)], and the resonances in $\rho_b$ [see lines $\mathsf{B}_{2}$, in Figs. 5 (c) and 5 (d))] become progressively wider.

\begin{figure*}[ht]
\centering
\includegraphics[width= 14 cm,height= 8 cm, scale= 1.0]
{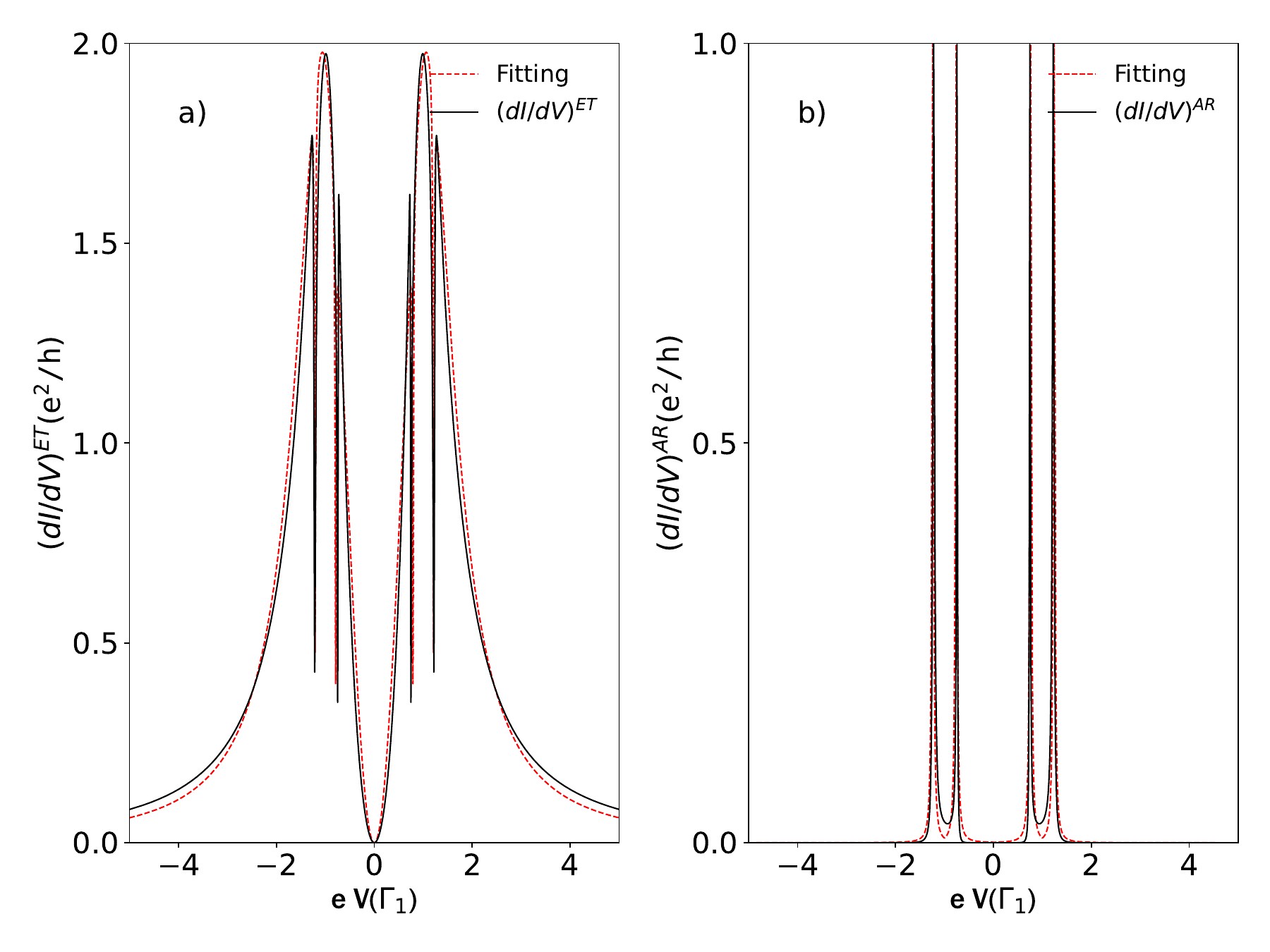}
\caption{\small{a) The differential conductance $(dI/dV)^{ET}$ (black solid line )  and the fitting (red dashed line)   as a function of the bias voltage $V$ when  $ t = 0.2$. b) The differential conductance $(dI/dV)^{AR}$ (black solid line ) and the fitting (red dashed line) as a function of the bias voltage $V$ when  $ t = 0.2$. Other parameters are chosen as $\Delta = 5 \Gamma_1$ , $k_B T = 0 \Gamma_1$, $\Gamma_1=\Gamma_2=\Gamma_S$  and  $\epsilon_a = \epsilon_b =-U/2$}.} 
\label{figure8}
\end{figure*}

\begin{figure*}[ht]
\centering
\includegraphics[width= 14 cm,height= 8 cm, scale= 1.0]
{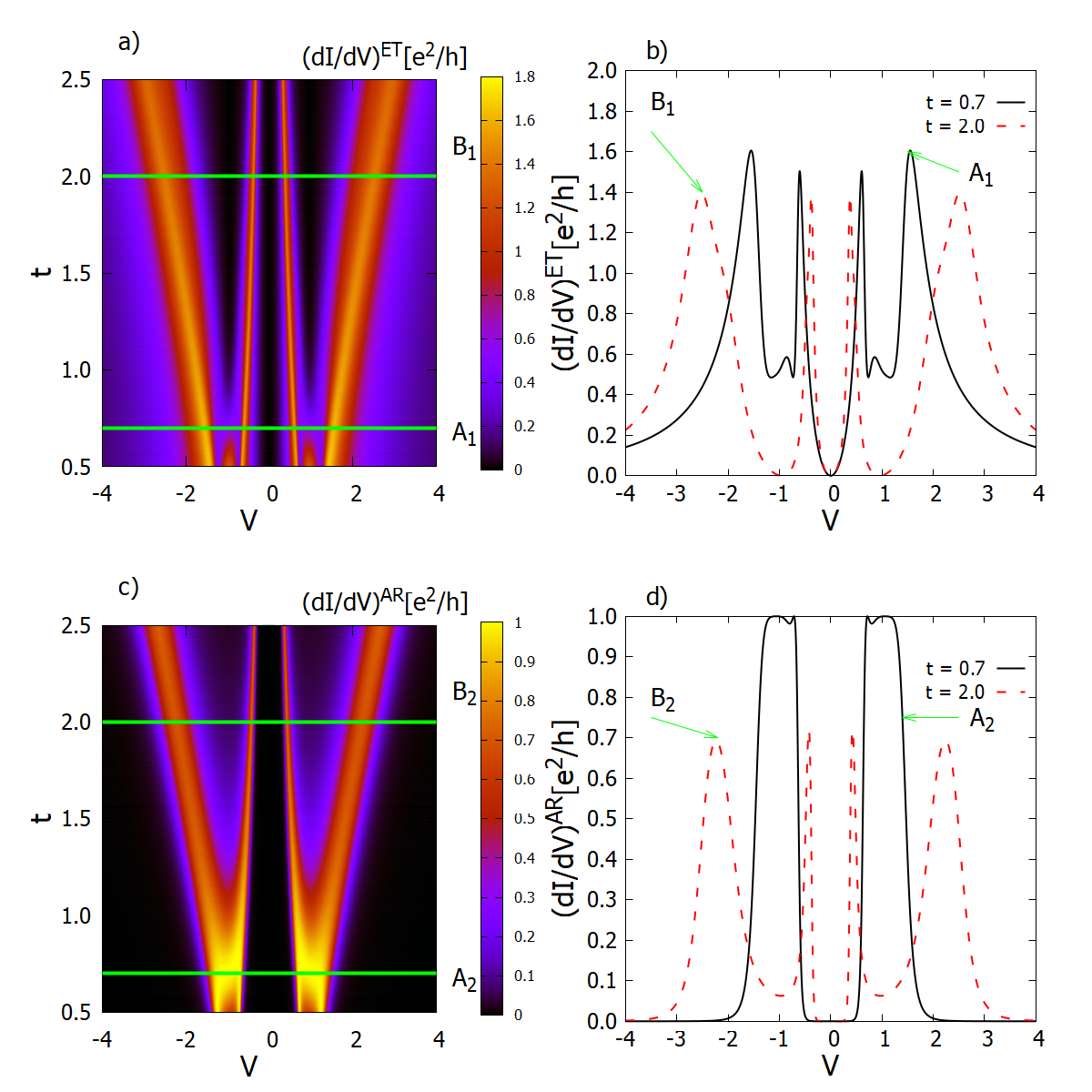}
\caption{\small{Left panel: Contour plot of 
$(dI/dV)^{ET}$ and $(dI/dV)^{AR}$ in terms  of $t$ and $V$, figures a) and c), respectively ; Right panel:$(dI/dV)^{ET}$ and $(dI/dV)^{AR}$  in terms of the bias voltage when $t = 0.7$ (solid line)  or $t = 2 $(dashed line), figures b) and d), respectively. Their location at the contour plot are indicated by the horizontal lines labeled by  A\textsubscript{1} and A\textsubscript{2}  for $t = 0.7$  and by B\textsubscript{1} and B\textsubscript{2}  for $t = 2.0$, respectively. Other parameters are chosen as $U = 2$, $k_B T = 0 \Gamma_1$, $\Gamma_1=\Gamma_2 = \Gamma_S$ and  $\epsilon_{a} = \epsilon_{b} =-U/2$  }}
\label{figure9}
\end{figure*}

In Fig. (\ref{figure6}) we study the role of the superconductor on  $(dI/dV)^{ET}$ by varying the coupling ratio, $r=\Gamma_{S}/\Gamma_{1}$, and $\Delta$, the superconductor gap. While $r$ measures the coupling strength between $QD_{b}$ and $S$, $\Delta$ carries the information of the superconductor correlations. The differential conductance $(dI/dV)^{ET}$  and $(dI/dV)^{AR}$, as a  function of bias voltage, are shown in the left and right panel of the Fig. (\ref{figure6}), respectively, for $r$ changing from 0.02 to 1.0.  For $\Delta=0$, the blue-dotted lines represent $(dI/dV)^{ET}$. In the first place, we can note  when $r = 0.02$,  i.e., when the DQD can be considered decoupled from the third lead, the difference conductance ET shows two deep Fano antiresonances, which reveals the existence of quantum interference due to the effect of the additional channel opened by the second QD coupled to the system. As the value of $r$ increases, the Fano antiresonances in the differential conductance ET disappear when the third lead is normal ($\Delta = 0$), due to the hybridization of the discrete state of $QD_{b}$ with the continuum spectrum of $S$ in its normal state. On the contrary, when $\Delta \ne 0$ and $r \ne 0$, the lead $S$ becomes a superconductor and the energy gap is revealed at the Fermi level with the presence of the ABSs, and two pairs of Fano antiresonances appear in the differential conductance ET, corresponding to the Fano resonances appearing in the differential conductance AR for the same value of $V$, which is a clear manifestation of quantum interference. For example, when $r=0.02$, the coupling to $S$ is too small, and  $(dI/dV)^{ET}$ exhibits the Fano anti-resonance pattern. In contrast, for $r\geq 0.50$, the ABSs become pronounced enough to change the shape of $(dI/dV)^{ET}$; this can be seen from the  $(dI/dV)^{AR}$ shown by the red curves in Fig. \ref{figure6}. As in the non-interacting case, the resonances of  $(dI/dV)^{AR}$ correspond to two anti-resonances in  $(dI/dV)^{ET}$ curves, equidistant from $\pm U/2$ with a central peak located at $\pm U/2$. In contrast to the normal state, this effect is robust concerning the increase of $r$, being preserved for all values of $r$. Actually, the main effect of $r$, in this case, is to displace the ABSs in the bias axis. Despite similar effects pointed out in the literature\cite{CalleA,CalleAM,J.Baranski}, here it is clear that such an interference pattern is robust against the Coulomb correlations, which, in general, have a detrimental effect on Andreev interference.

Finally, Fig.  \ref{figure7} displays in the left panel the electron-tunneling differential conductance($(dI/dV)^{ET}$) (black solid line)  and in the right panel the Andreev differential conductance ($(dI/dV)^{AR}$) (red solid line) for different values of intra-dot Coulomb interaction $U$.  The dashed line in the left panel also corresponds to the $(dI/dV)^{ET}$  for $\Delta = 0$, i.e., for a system with three normal contacts.  We can see in Fig.  \ref{figure7} that when $t = 0.2$, $U =0$ and $\Delta = 0$, the differential conductance ET has the form of a Lorenzian centered on the zero bias voltage and does not present anti-resonances. On the contrary, when  $U =0$  and $\Delta \ne 0$, the differential conductance ET shows two Fano antiresonances, which coincide with the two resonances in the differential conductance AR. Conversely, when $U \ne 0$ and $\Delta \ne 0$, the two antiresonances in the differential conductance ET and the two resonances in the differential conductance AR are subdivided on either side of the zero-bias with the symmetry point located at $\pm U/2$, by the effect of the Coulomb interaction. The separation between the resonances in the differential conductance ET and AR grows as $U$ increases.

One way to better comprehend the tunneling mechanism in different transmission processes is to analyze the differential conductance $(dI/dV)^{ET}$. It can be expressed as a convolution of two Breit-Wigner and two Fano line shapes, as shown in Figure 8a.

\begin{multline}
\frac{dI^{ET}}{dV} \approx \frac{|\Tilde{V}_1 + q|^2}{|\Tilde{V}_1|^2 + 1}  \left(  \frac{1}{\epsilon_1^2 + 1} +  \frac{1}{\epsilon_2^2 + 1} \right)\frac{|\Tilde{V}_2|^2}{|\Tilde{V}_2|^2 + 1} ,
\label{FitFano}
\end{multline}
where  $ \Tilde{V}_1 = (||V| - U/2| - \Gamma_S/4)/\eta $ with $\eta = t^2/(\Gamma_S) $, $\Tilde{V}_2 = |V|/(\Gamma_{L1}/2) $ , $\epsilon_1=\frac{|V|- \epsilon_d}{(\Gamma_{L1}/2)}$,$\epsilon_2=\frac{|V|- (\epsilon_d + U)}{(\Gamma_{L1}/2)}$and  $q = q_r + i\, q_i$.  The complex Fano parameter, $q$, and the coupling value, $\Gamma_{L1}$, are relevant to understanding the differential conductance. Based on the above equation, the conductance has two resonances at $V= \epsilon_d$ and $V= \epsilon_d + U$. Each resonance has two dips at $V= \epsilon_d \pm \Gamma_S/4$ and  $V= (\epsilon_d + U) \pm \Gamma_S/4$, with $\epsilon_d = -U/2$. Additionally, a single antiresonance appears at $V=0$.  The interference of electrons through different tunneling paths, including Andreev bound-states, explains the Fano effect observed in our results. Due to the presence of Andreev bound states, the electron suffers a change in its phase, and this is reflected in a complex $q$-parameter as we see in Eq.(\ref{FitFano}).
In addition, the Andreev differential conductance, $(dI/dV)^{AR}$, can be expressed as two Breit-Wigner line shapes (shown in Figure 8b).

\begin{equation}
\frac{dI^{AR}}{dV} \approx  \left(  \frac{1}{\epsilon_3^2 + 1} +  \frac{1}{\epsilon_4^2 + 1} \right),
\end{equation}
where $\epsilon_3=\frac{|V- \epsilon_d|- \Gamma_S/4}{(t^2/\Gamma_{S})}$ and $\epsilon_4=\frac{|V- (\epsilon_d + U)| - \Gamma_S/4}{(t^2/\Gamma_{S})}$.
The above equation shows that the Andreev differential conductance has four resonances at $V= \epsilon_d \pm \Gamma_S/4$ and  $V= (\epsilon_d + U) \pm \Gamma_S/4$, with $\epsilon_d = -U/2$,  which coincide exactly with the dips of the differential conductance ET mentioned above. From the fits, we can conclude that the Fano-Andreev effect is robust in the presence of Coulomb interaction. Additionally, in the interferometric regime, Andreev transmission is a resonant tunneling process through Andreev-bound states, which causes the Fano effect in normal transmission.


\begin{figure*}[ht]
\centering
\includegraphics[width= 14 cm,height= 8 cm, scale= 1.0]
{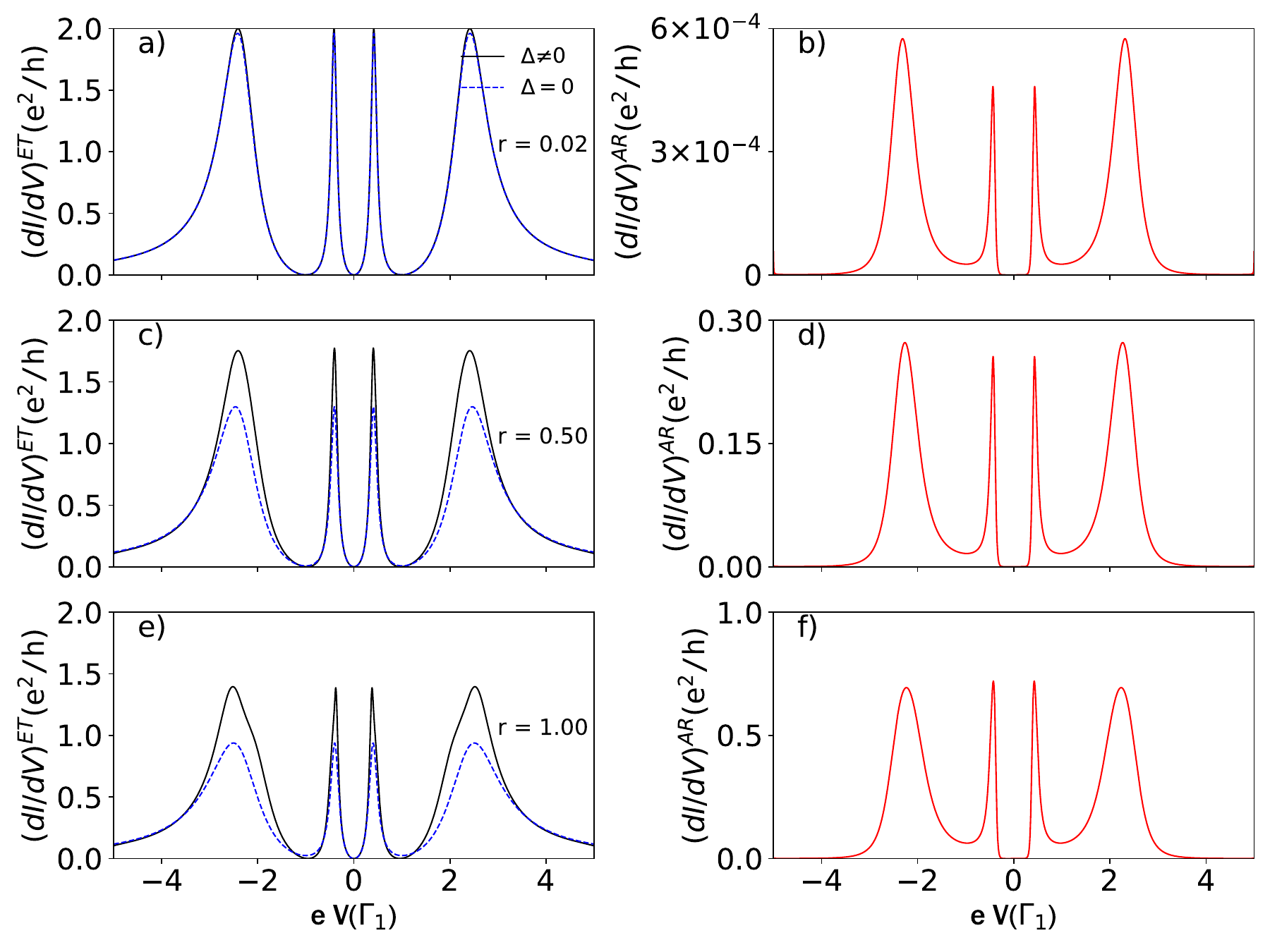}
\caption{\small{In the left panel, the differential conductance $(dI/dV)^{ET}$ (black solid line) as a function of the bias voltage $V$ and in the right panel $(dI/dV)^{AR}$ (red solid line )  as a function of the bias voltage $V$ when   $ r = 1$ for a) and b), $ r = 2$ for c) and d), and $ r = 4$ for e) and f). The blue dashed line in the left panel corresponds to the $(dI/dV)^{ET}$  for $\Delta = 0$, i.e., for a system with three normal contacts. Other parameters are chosen as $t = 2 \Gamma_1$ , $\Delta = 5 \Gamma_1$ , $U = 2 \Gamma_1$, $k_B T = 0 \Gamma_1$, $\Gamma_1=\Gamma_2$  and  $\epsilon_a = \epsilon_b = -U/2$}.} 
\label{figure10}
\end{figure*}

\begin{figure*}[ht]
\centering
\includegraphics[width= 14 cm,height= 8 cm, scale= 1.0]
{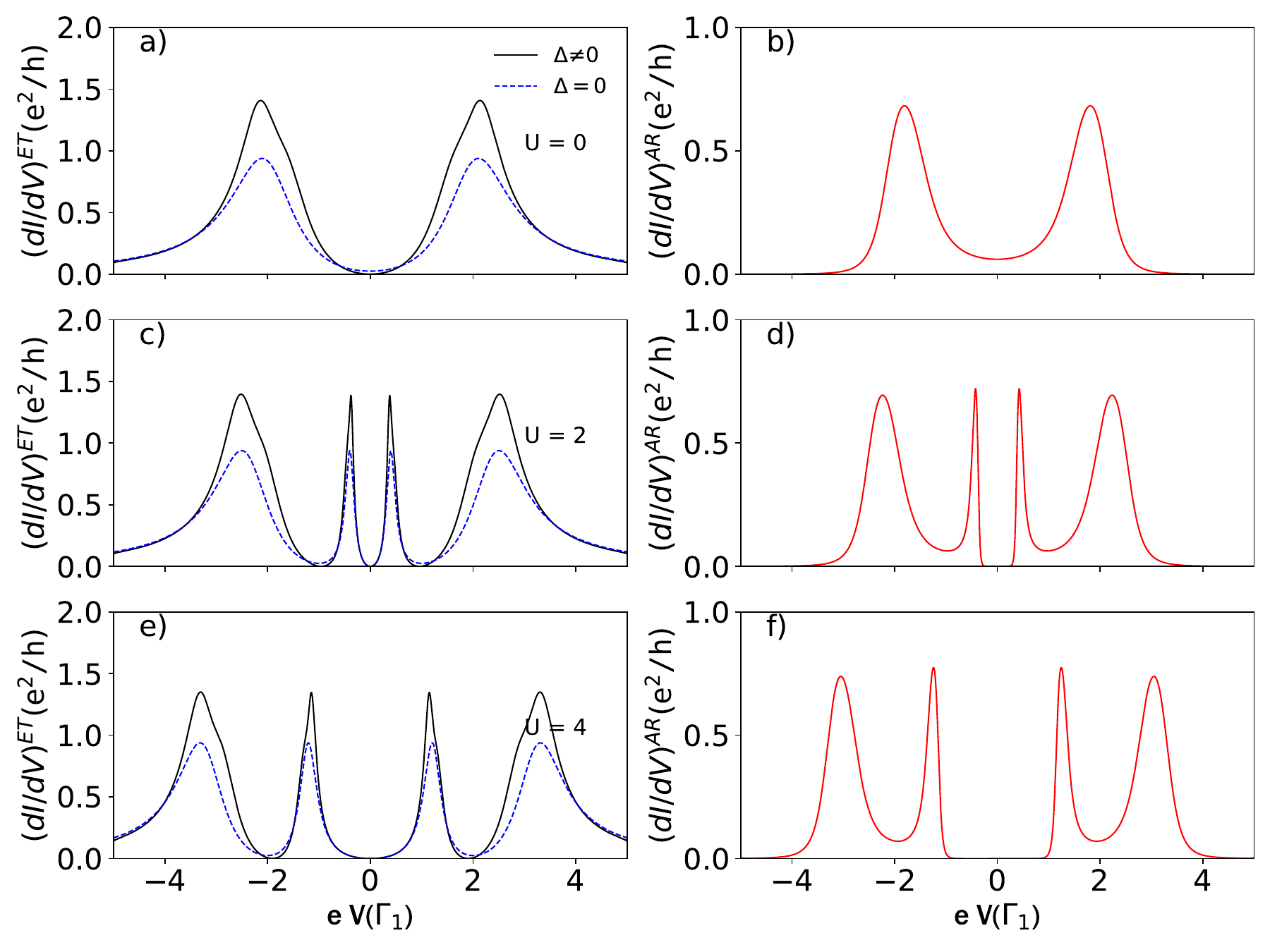}
\caption{\small{In the Left panel, the differential conductance $(dI/dV)^{ET}$ (black solid line )  as a function of the bias voltage $V$ and in the right panel $(dI/dV)^{DAR}$ (red solid line ) as a function of the bias voltage $V$ when  $ U = 0$ for a) and b), $ U = 2$ for c) and d), and $ U = 4$ for e) and f). The blue dashed line in the left panel corresponds to the $(dI/dV)^{ET}$  for $\Delta = 0$, i.e., for a system with three normal contacts. Other parameters are chosen as $t = 2 \Gamma_1$ , $\Delta = 5 \Gamma_1$ , $k_B T = 0 \Gamma_1$, $\Gamma_1=\Gamma_2=\Gamma_S$  and  $\epsilon_a = \epsilon_b =-U/2$}.} 
\label{figure11}
\end{figure*}

\subsection{Molecular regime}

\begin{figure*}[ht]
\centering
\includegraphics[width= 14 cm,height= 8 cm, scale= 1.0]
{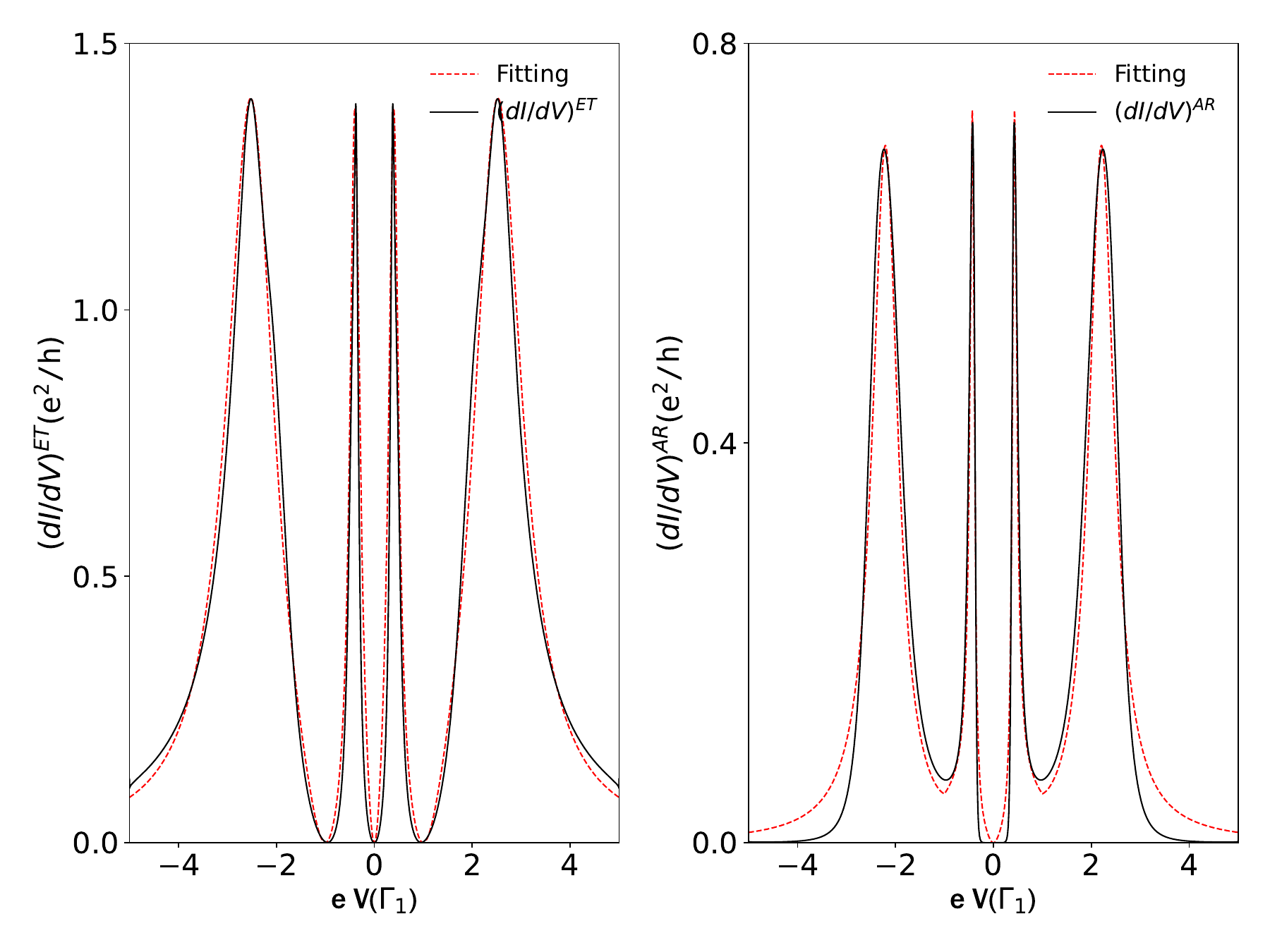}
\caption{\small{a) The differential conductance $(dI/dV)^{ET}$ (black solid line )  and the fitting (red dashed line)  as a function of the bias voltage $V$ when  $ t = 2$. b) The differential conductance $(dI/dV)^{AR}$ (black solid line ) and the fitting (red dashed line)  as a function of the bias voltage $V$ when  $ t = 0.2$. Other parameters are chosen as $\Delta = 5 \Gamma_1$ , $k_B T = 0 \Gamma_1$, $\Gamma_1=\Gamma_2=\Gamma_S$  and  $\epsilon_a = \epsilon_b =-U/2$}.} 
\label{figure12}
\end{figure*}

In this subsection, we analyze the molecular regime. The contour plots in Fig. \ref{figure9} a-c display the behavior of $(dI/dV)^{ET}$ and $(dI/dV)^{AR}$  as functions of $t$ and $V$, in the range $0.5 \Gamma_1 \leq t \leq 2.5 \Gamma_1$, are shown in the contour plots of Figs. \ref{figure9}(a) and \ref{figure9}(c), respectively. The other parameters remain the same as the ones used in Figs. \ref{figure4} and \ref{figure5}. 
This figure displays the progressive vanishing of the antiresonances in the differential conductance ET as $t$ increases.  As seen in the black solid line in Fig. 9 a or line $\mathsf{A}_{1}$ in 9 b, when $t = 0.7$, the antiresonances in the differential conductance ET have almost disappeared. Therefore, we can consider that this value of $t$ is the threshold value that defines the transition between the interferometric and molecular regimes. On the other hand, when $t > 0.7$, specifically at $t = 2$ (see line B1 in Fig. 9 a, or red dashed line in Fig. 9 b), the differential conductance ET becomes zero at $\pm U/2$, with two asymmetrically located resonances around this point: two lateral resonances moving away from the $eV=0$, and two central resonances approaching each other towards the point of zero bias voltage. On the other hand, the differential conductance AR exhibits similar behavior. At $t=0.7$ (indicated by the black solid line on Fig. 9d or line A2 on Fig. 9c), the double peak structure has almost vanished only to reappear again for $t>1$ (as shown by the red dashed line in Fig. 9d or line B2 in Fig. 9c). However, now the resonances in the differential Andreev conductance align closely with those in the differential ET conductance (as seen in line B2 in Fig. 9c).

Now, we study the effect of the coupling ratio, $r$, on the differential conductance ET and AR in the molecular regime ($t \geq \Gamma_1$). Fig. \ref{figure10} displays in the left panel the electron-tunneling differential conductance($(dI/dV)^{ET}$) (black solid line), and in the right panel, the Andreev differential conductance ($(dI/dV)$) (red solid line), for different values of rate coupling $r$  for $\Delta = 5$. The blue dashed line in the left panel also corresponds to the $(dI/dV)^{ET}$  for $\Delta = 0$, i.e., for a system with three normal contacts. In the first place, we can note that by changing $r$ from 0.05 to 1, the shape of the differential conductance curve ET does not change significantly when the lead S is in its normal state ($\Delta = 0$) or when it is in its superconducting state ($\Delta \neq 0$). The only noticeable effect on the differential conductance of the presence of lead S in its superconducting state is the decrease in the height of the resonances and the progressive splitting of the lateral peaks. On the other hand, the differential conductance AR presents two resonances on each side of the zero bias voltage, asymmetrically located around $\pm U/2$, which are located approximately around the same bias voltage values as the resonances in the differential conductance ET, and these resonances turn out to be visible even for small values of $r$. However, it is possible to appreciate a remarkable reduction in the height of the AR differential conductance resonances when we are in the molecular regime for the interferometric regime (see Fig. 10 b, d, and f).

Fig.  \ref{figure11} displays in the left panel the electron-tunneling differential conductance ET (black solid line)  and in the right panel the Andreev differential conductance (red solid line), in the molecular regime, for different values of intra-dot Coulomb interaction $U$.  The blue dashed line in the left panel also corresponds to the $(dI/dV)^{ET}$  for $\Delta = 0$, i.e., for a system with three normal contacts.  We can see in Fig.  \ref{figure11} that when $t = 2$, $U =0$, in contrast to the interferometric regime,  the differential conductance ET and AR have the form of two Lorenzian curves, localized close to $\pm t$ with the center of symmetry located at zero bias voltage,  irrespective of whether the lead S is in its normal state ($\Delta = 0$) or its superconducting state ($\Delta \neq 0$), and unlike in the interferometric case, there are no antiresonances present in $(dI/dV)^{ET}$ when ($\Delta \neq 0$).  On the other hand, when $U$ is non-zero, whether $\Delta = 0$ or $\Delta \neq 0$,   the resonances in $(dI/dV)^{ET}$ and $(dI/dV)^{AR}$  are doubled, and are located approximately around the bias voltage $ \epsilon_d $ and $ \epsilon_d + U$, by the effect of the Coulomb interaction. The separation between the resonances in the differential conductance ET and AR grows as U increases. 

The differential conductance $(dI/dV)^{ET}$ can be represented by combining two Breit-Wigner and two Fano line shapes, in the range $0.7 \leq t \leq 2$ and $r \leq 1$,  as shown in Figure 12a.

\begin{equation}
\frac{dI^{ET}}{dV} \approx \frac{\epsilon_1^2 }{\epsilon_1^2 + 1}  \Big( \frac{1}{\omega_{1}^2 + 1} + \frac{1}{\omega_{2}^2 + 1} \Big) \frac{\epsilon_2^2 }{\epsilon_2^2 + 1} 
\end{equation}
where  $\omega_{1} = ||V| - \gamma_1|/\eta_{1} $, $\omega_{2} = (||V| -\gamma_2|)/\eta_2 $, $\epsilon_{1} = ||V| - U/2|/(\Gamma_{1}/2)$, $\epsilon_{2} = |V|/\Gamma_1$.  Where $\gamma_i = (-\epsilon_d \mp t/2 - U/2 + (1/2)\sqrt{t^2 + U^2 })$ with $i=1,2$ ( in the molecular limit $((t/\Gamma_{L1})> 1)$ and at the electron-hole symmetry point) and on other hand,$\eta_1$ and $\eta_2$  are fitting parameters.  Note that from the above expression, the differential conductance vanishes at $V=\pm U/2$ and $V=0$. The $q$-parameter in molecular regimes is real, unlike in interferometric regimes. Besides, the differential conductance shows resonances at $V \approx \pm \gamma_1$ and $V \approx \pm (\gamma_2)$. Additionally, it is interesting to note that by making a taylor approximation with respect to $t/U$ , the values of $\gamma_1$ and $\gamma_2$ become $\gamma_1= U^2/(4 t)$ and $\gamma_2=t + U^2/(4 t)$ , respectively. That is, when $t>>U$ the central peaks tend to approach $V=0$, and the lateral peaks tend to lie at $V=t$. 

In addition,  the equation for $(dI/dV)^{AR}$ may be written as the superposition of two  Breit-Wigner and one Fano line shapes. [Fig. \ref{figure12}(b)]:

\begin{equation}
\frac{dI^{AR}}{dV} \approx  \frac{\xi^2}{  \xi^2 + 1   }   (\frac{1}{\nu_{1}^2 + 1} + \frac{1}{\nu_{2}^2 + 1})
\end{equation}
where  $\nu_{1} = ||V| - \gamma_1|/(\bar{\xi}_{1}) $ , $\nu_{2} = (||V| - \gamma_2|)/(\bar{\xi}_2) $, and $\xi = |V|/(\bar{\xi}_3) $. Here, as before,$\gamma_i = (-\epsilon_d \mp t/2 - U/2 + (1/2)\sqrt{t^2 + U^2 })$ with $i=1,2$ ( in the molecular limit $((t/\Gamma_{L1})> 1)$ and at the electron-hole symmetry point) and on other hand,$\bar{\xi}_1$,$\bar{\xi}_2$ and $\bar{\xi}_3$  are fitting parameters. 
 
 The fitting presented above offers a reasonable understanding of the shape of the Andreev differential conductance shown in Figure 11. The Andreev states undergo a split caused by $t$ and as a result of this tunneling coupling, the Andreev bound states acquire widths that manifest in the Andreev differential conductance.


\section{Summary}\label{secconclu}


We have investigated the electronic transport properties of a T-shaped double quantum dot (DQD) system in the Coulomb blockade regime under non-equilibrium conditions. We employed a non-equilibrium Green's function calculation method and the equation of motion approach using the Hubbard-I approximation to do this. Our results suggest superconducting interference effects on transport between normal leads, which can be identified as Fano-like anti-resonances in the QD transmission spectrum.

We have identified two distinct regimes, the interferometric and molecular regimes. In both regimes, the differential conductance (ET) can be expressed as a convolution of Fano and Breit Wigner lines shape. However, in the interferometric regime, the Fano line shapes are centered on the energies of the Andreev bound states with a finite complex $q$-parameter, while in the molecular regime, the $q$-parameter takes real values. On the other hand, the Andreev reflection exhibits maxima that correspond to the Andreev bound states. Therefore, we can conclude that the interference effect is robust against Coulomb correlations and can be experimentally probed under non-equilibrium conditions.


\section*{Acknowledgements}

This work has been partially supported by  Universidad Santa Mar\' ia  Grant USM-DGIIP PI-LI 1925 and 1919 and FONDECYT Grant 1201876.




\bibliographystyle{unsrt}
\bibliography{bibliography_paper2}

\end{document}